\pageno=1                                      
\input psfig.tex

\def\Ha{H$\alpha$}

%
%
%
\font\ninerm=cmr9
\font\eightrm=cmr8
\font\sixrm=cmr6
\font\ninei=cmmi9
\font\eighti=cmmi8
\font\sixi=cmmi6
\skewchar\ninei='177 \skewchar\eighti='177 \skewchar\sixi='177
\font\ninesy=cmsy9
\font\eightsy=cmsy8
\font\sixsy=cmsy6
\skewchar\ninesy='60 \skewchar\eightsy='60 \skewchar\sixsy='60

\font\ninebf=cmbx9
\font\eightbf=cmbx8
\font\sixbf=cmbx6
\font\ninett=cmtt9
\font\eighttt=cmtt8
\hyphenchar\tentt=-1 
\hyphenchar\ninett=-1
\hyphenchar\eighttt=-1
\font\ninesl=cmsl9
\font\eightsl=cmsl8
\font\nineit=cmti9
\font\eightit=cmti8
\newskip\ttglue
\def\tenpoint{\def\rm{\fam0\tenrm}%
  \textfont0=\tenrm \scriptfont0=\sevenrm \scriptscriptfont0=\fiverm
  \textfont1=\teni \scriptfont1=\seveni \scriptscriptfont1=\fivei
  \textfont2=\tensy \scriptfont2=\sevensy \scriptscriptfont2=\fivesy
  \textfont3=\tenex \scriptfont3=\tenex \scriptscriptfont3=\tenex
  \def\it{\fam\itfam\tenit}%
  \textfont\itfam=\tenit
  \def\sl{\fam\slfam\tensl}%
  \textfont\slfam=\tensl
  \def\bf{\fam\bffam\tenbf}%
  \textfont\bffam=\tenbf \scriptfont\bffam=\sevenbf
   \scriptscriptfont\bffam=\fivebf
  \def\tt{\fam\ttfam\tentt}%
  \textfont\ttfam=\tentt
  \tt \ttglue=.5em plus.25em minus.15em
  \normalbaselineskip=12pt
  \let\sc=\eightrm
  \let\big=\tenbig
  \setbox\strutbox=\hbox{\vrule height8.5pt depth3.5pt width0pt}%
  \normalbaselines\rm}
\def\ninepoint{\def\rm{\fam0\ninerm}%
  \textfont0=\ninerm \scriptfont0=\sixrm \scriptscriptfont0=\fiverm
  \textfont1=\ninei \scriptfont1=\sixi \scriptscriptfont1=\fivei
  \textfont2=\ninesy \scriptfont2=\sixsy \scriptscriptfont2=\fivesy
  \textfont3=\tenex \scriptfont3=\tenex \scriptscriptfont3=\tenex
  \def\it{\fam\itfam\nineit}%
  \textfont\itfam=\nineit
  \def\sl{\fam\slfam\ninesl}%
  \textfont\slfam=\ninesl
  \def\bf{\fam\bffam\ninebf}%
  \textfont\bffam=\ninebf \scriptfont\bffam=\sixbf
   \scriptscriptfont\bffam=\fivebf
  \def\tt{\fam\ttfam\ninett}%
  \textfont\ttfam=\ninett
  \tt \ttglue=.5em plus.25em minus.15em
  \normalbaselineskip=10pt 
  \let\sc=\sevenrm
  \let\big=\ninebig
  \setbox\strutbox=\hbox{\vrule height8pt depth3pt width0pt}%
  \normalbaselines\rm}
\def\eightpoint{\def\rm{\fam0\eightrm}%
  \textfont0=\eightrm \scriptfont0=\sixrm \scriptscriptfont0=\fiverm
  \textfont1=\eighti \scriptfont1=\sixi \scriptscriptfont1=\fivei
  \textfont2=\eightsy \scriptfont2=\sixsy \scriptscriptfont2=\fivesy
  \textfont3=\tenex \scriptfont3=\tenex \scriptscriptfont3=\tenex
  \def\it{\fam\itfam\eightit}%
  \textfont\itfam=\eightit
  \def\sl{\fam\slfam\eightsl}%
  \textfont\slfam=\eightsl
  \def\bf{\fam\bffam\eightbf}%
  \textfont\bffam=\eightbf \scriptfont\bffam=\sixbf
   \scriptscriptfont\bffam=\fivebf
  \def\tt{\fam\ttfam\eighttt}%
  \textfont\ttfam=\eighttt
  \tt \ttglue=.5em plus.25em minus.15em
  \normalbaselineskip=9pt
  \let\sc=\sixrm
  \let\big=\eightbig
  \setbox\strutbox=\hbox{\vrule height7pt depth2pt width0pt}%
  \normalbaselines\rm}
%
\def\headtype{\ninepoint}                 
\def\abstracttype{\ninepoint}             
\def\captiontype{\ninepoint}              
\def\footnotetype{\ninepoint}             
\def\refit{\it}                           
\font\chaptitle=cmr10 at 11pt             
\rm                                       

%
%
\parindent=0.25in                         
\parskip=0pt                              
\baselineskip=12pt                        
\hsize=4.25truein                         
\vsize=7.445truein                        
\hoffset=1in                              
\voffset=-0.5in                           

\newskip\sectionskipamount                
\newskip\aftermainskipamount              
\newskip\subsecskipamount                 
\newskip\firstpageskipamount              
\newskip\capskipamount                    
\newskip\ackskipamount                    
\sectionskipamount=0.2in plus 0.09in
\aftermainskipamount=6pt plus 6pt         
\subsecskipamount=0.1in plus 0.04in
\firstpageskipamount=3pc
\capskipamount=0.1in
\ackskipamount=0.15in
\def\sectionskip{\vskip\sectionskipamount}
\def\aftermainskip{\vskip\aftermainskipamount}
\def\subsecskip{\vskip\subsecskipamount} 
\def\firstpageskip{\vskip\firstpageskipamount}
\def\capskip{\hskip\capskipamount}

%
%
\nopagenumbers                            
\newcount\firstpageno                     
\firstpageno=\pageno                      
\newcount\chapno                          

\def\rightheadline{\headtype\phantom{\folio}\hfil\runningtitletext\hfil\folio}
\def\leftheadline{\headtype\folio\hfil\runningauthortext\hfil\phantom{\folio}}
\headline={\ifnum\pageno=\firstpageno\hfil
           \else
              \ifdim\ht\topins=\vsize           
                 \ifdim\dp\topins=1sp \hfil     
                 \else
                     \ifodd\pageno\rightheadline\else\leftheadline\fi
                 \fi
              \else
                 \ifodd\pageno\rightheadline\else\leftheadline\fi
              \fi
           \fi}

\def\bottomnumber{\hss\tenrm[\folio]\hss}
\footline={\ifnum\pageno=\firstpageno\bottomnumber\else\hfil\fi}

%
%
%
%
\outer\def\mainsection#1
    {\vskip 0pt plus\smallskipamount\sectionskip
     \message{#1}\vbox{\noindent{\bf#1}}\nobreak\aftermainskip\noindent}
 
\outer\def\subsection#1
    {\vskip 0pt plus\smallskipamount\subsecskip
     \message{#1}\vbox{\noindent{\bf#1}}\nobreak\smallskip\nobreak\noindent}
 
\def\backup{\nobreak\vskip-\baselineskip\nobreak\vskip-\subsecskipamount\nobreak
}

\def\title#1{{\chaptitle\leftline{#1}}}
\def\name#1{\leftline{#1}}
\def\affiliation#1{\leftline{\it #1}}
\def\abstract#1{{\abstracttype \noindent #1 \smallskip\vskip .1in}}
\def\ref{\noindent \parshape2 0truein 4.25truein 0.25truein 4truein}
\def\caption{\noindent \captiontype
             \parshape=2 0truein 4.25truein .125truein 4.125truein}

\def\footnote#1{\edef\fspafac{\spacefactor\the\spacefactor}#1\fspafac
      \insert\footins\bgroup\footnotetype
      \interlinepenalty100 \let\par=\endgraf
        \leftskip=0pt \rightskip=0pt
        \splittopskip=10pt plus 1pt minus 1pt \floatingpenalty=20000
        \textindent{#1}\bgroup\strut\aftergroup\strut\egroup\let\next}
\skip\footins=12pt plus 2pt minus 4pt 
\dimen\footins=30pc 

%
%

\def\@{\spacefactor 1000}

\def\,{\pcomma} 
\def\pcomma{\relax\ifmmode\mskip\thinmuskip\else\thinspace\fi}

\def\oversim#1#2{\lower0.5ex\vbox{\baselineskip=0pt\lineskip=0.2ex
     \ialign{$\mathsurround=0pt #1\hfil##\hfil$\crcr#2\crcr\sim\crcr}}}

\def\runningtitletext{Brown Dwarfs}
\def\runningauthortext{Oppenheimer et al.}

\null
\firstpageskip

{\baselineskip=14pt
\title{BROWN DWARFS}
}

\vskip .3truein
\name{B. R. OPPENHEIMER, S. R. KULKARNI}
\affiliation{Palomar Observatory, California Institute of Technology}
\vskip .1truein
\leftline{and}
\vskip .1truein
\name{J. R. STAUFFER}
\affiliation{Harvard--Smithsonian Center for Astrophysics}
\vskip .3truein

\abstract{After a discussion of the physical processes in brown 
dwarfs, we present a complete, precise definition of brown dwarfs and
of planets inspired by the internal physics of objects between 0.1 and
0.001 $M_{\odot}$.  We discuss observational techniques for
characterizing low-luminosity objects as brown dwarfs, including the
use of the lithium test and cooling curves.  A brief history of the
search for brown dwarfs leads to a detailed review of known isolated
brown dwarfs with emphasis on those in the Pleiades star cluster.  We
also discuss brown dwarf companions to nearby stars, paying particular
attention to Gliese 229B, the only known cool brown dwarf.}

\mainsection{I.  WHAT IS A BROWN DWARF?}
\backup

A main sequence star is to a candle as a brown dwarf is to a hot poker
recently removed from the fire.  Stars and brown dwarfs, although they
form in the same manner, out of the fragmentation and gravitational
collapse of an interstellar gas cloud, are fundamentally different
because a star has a long-lived internal source of energy: the fusion
of hydrogen into helium.  Thus, like a candle, a star will burn
constantly until its fuel source is exhausted.  A brown dwarf's core
temperature is insufficient to sustain the fusion reactions common to
all main sequence stars.  Thus brown dwarfs cool as they age.  Cooling
is perhaps the single most important salient feature of brown dwarfs,
but an understanding of their definition and their observational
properties requires a review of basic stellar physics.

\subsection{A. Internal Physics}

In stellar cores, nuclear fusion acts as a strict thermostat
maintaining temperatures very close to the nuclear fusion temperature,
$T_{\rm nucl} = 3 \times 10^6$ K, the temperature above which hydrogen
fusion becomes possible.  In the stellar core, the velocities of the
protons obey a Maxwell-Boltzmann distribution.  However, the average
energy of one of these protons is only $kT_{\rm nucl} = 8.6 \times
10^{-8}T$ keV $\sim 0.1$ keV, where $k$ is the Boltzmann constant.  In
contrast, the Coulomb repulsion between these protons is on the order
of MeV.  Despite this enormous difference in energies, fusion is
possible because of quantum mechanical tunneling.  The nuclear
reaction rate is governed by the proton pair energy, $E$, at the high
energy tail of the Maxwell-Boltzmann distribution, which scales as
$\exp(-E/kT)$, and the nuclear cross section due to quantum mechanical
tunneling through the Coulomb repulsion, which scales as $\exp
(-E^{-1/2})$.  The product of these two factors defines a sharp peak,
the Gamov peak, at a critical energy $E_{\rm crit}$.  $E_{\rm crit}$
is approximately 10 keV for the reactions in the pp chain, the most
basic form of hydrogen fusion.  Because $E_{\rm crit} >> kT$, the
nuclear reactions involve the tiny minority of protons in the high
velocity tail of the Maxwell-Boltzmann distribution.  In terms of
temperature, this reaction rate is proportional to $(T/T_{\rm
nucl})^n$ where $n \approx 10$ for temperatures near $T_{\rm nucl}$
and reactions in the pp chain.  The large value of $n$ ensures that
the core temperature is close to $T_{\rm nucl}$.

For the low-mass main sequence stars in which the above discussion
holds, the mass is roughly proportional to the radius.  This can be
shown with a simplified argument by appealing to the virial theorem.
In equilibrium, the thermal energy and the gravitation potential
energy are in balance: $GM^2/R \sim (M/m_{p}) kT_{\rm
nucl}$, where $m_p$ is the mass of the proton, $M$ is the mass of the
star, $R$ is its radius, and $G$ is the gravitational constant.
Therefore, $R
\propto M$.

If radius is proportional to mass, then the density, $\rho$, increases
with decreasing mass: $\rho \propto MR^{-3} \propto M^{-2}$.  At a
high enough density a new source of pressure becomes important.
Electrons, because they have half integral spins, must obey the Pauli
exclusion principle and are accordingly forbidden from occupying
identical quantum energy states.  This requires that electrons
successively fill up the lowest available energy states.  The
electrons in the higher energy levels contribute to degeneracy
pressure, because they cannot be forced into the filled, lower energy
states.  The degeneracy pressure, which scales as $\rho^{5/3}$,
becomes important when it approximately equals the ideal gas pressure:
$\rho T \propto\rho^{5/3}$.  Explicit calculation of this relation
shows that degeneracy pressure dominates when $\rho > 200$ g cm$^{-3}$
and $T < T_{nucl}$.  For the Sun, $\rho \approx 1$ gm cm$^{-3}$.
Using the scaling relations above, one finds that degeneracy pressure
becomes important for stars with $M < 0.1 M_{\odot}$.  Objects
supported primarily by some sort of degeneracy pressure are called
``compact.''

An examination of Fig.\ 1 (Burrows and Liebert 1993) demonstrates the
key elements described above.  This plot of mass versus radius shows
the main sequence (labeled ``M Dwarfs''), where $R \propto M$, and a
line which white dwarfs must obey because they are completely
supported by electron degeneracy pressure.  Thus, the energy density
of the degenerate electrons ($\propto \rho^{5/3}$) must match the
gravitational potential energy density ($GM^2 / R / R^3$).  In that
case, $R \propto M^{-1/3}$.  Note that the white dwarf sequence meets
the main sequence at about 0.1 $M_{\odot}$.  At this point, the main
sequence curve turns and remains at a roughly constant radius for all
the masses down to the mass of Jupiter.  This mass range, from about
0.1 $M_{\odot}$ to 0.001 $M_{\odot}$, has an essentially constant
radius because the degeneracy pressure leads to the slow function $R
\propto M^{-1/3}$ at the high mass end.  Then, at the low mass end the
Coulomb pressure, which is characterized by constant density ($\rho
\propto M / R^3$ which implies $R \propto M^{+1/3}$), begins to
dominate over degeneracy, the net result being approximately $R
\propto M^0$  (Burrows and Liebert 1993).

\smallskip

\centerline{\psfig{file=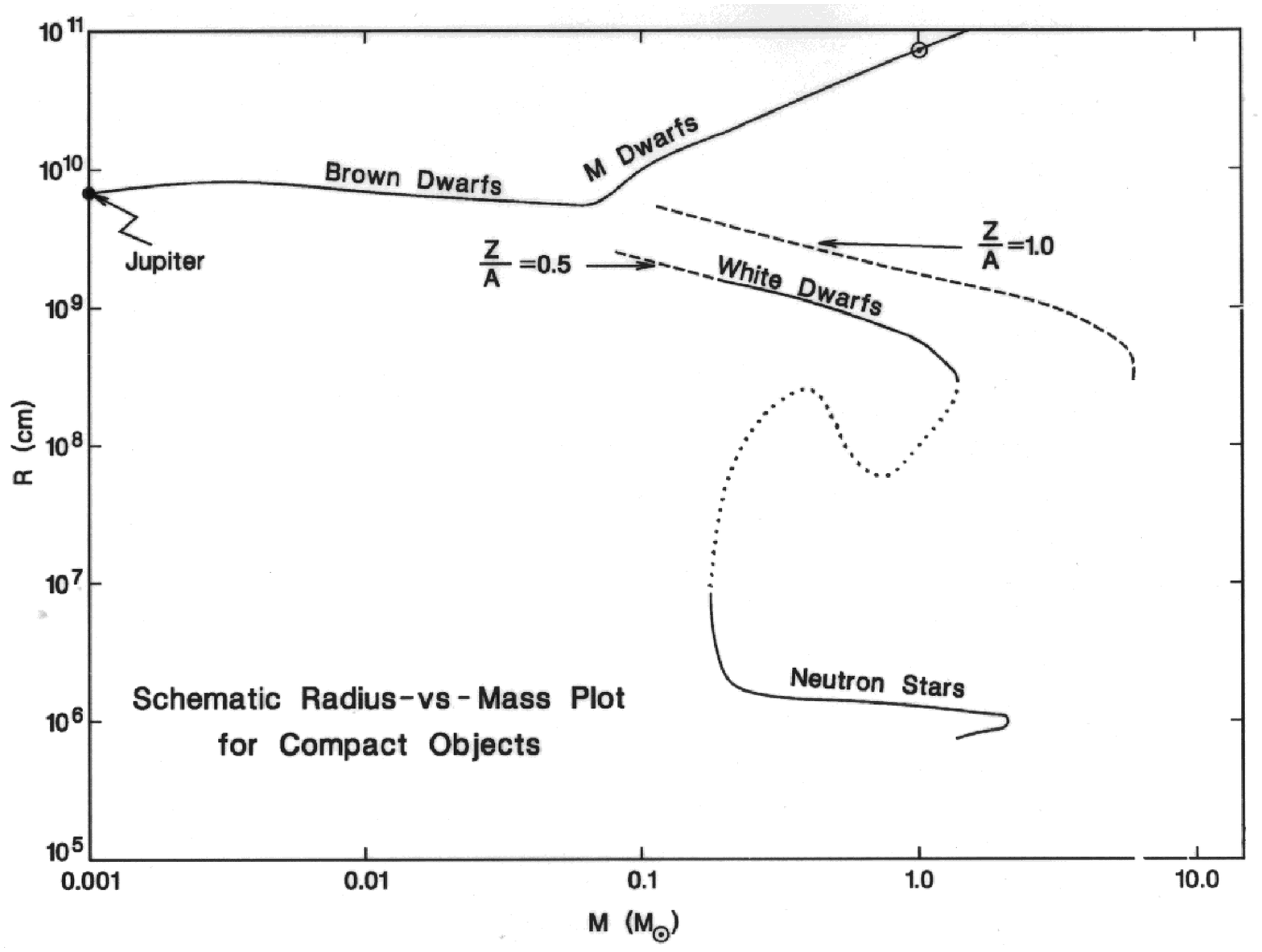,angle=270,width=4in}}

\noindent
{\caption{Fig.\ 1.\capskip Mass versus Radius.  This plot shows that
brown dwarfs have a roughly constant radius as a function of mass.
This is important because it makes $T_{\rm eff}$ a function only of
luminosity.  If $T_{\rm eff}$ is observable then an object can be
classified as a brown dwarf or star based on its $T_{\rm eff}$ or
luminosity.  (See Fig.\ 2.)  Courtesy A. Burrows. From Burrows and
Liebert (1993).}}

\smallskip

Kumar (1963) calculated the mass at which ``star-like'' objects could
be stable against gravitational collapse through electron degeneracy
pressure, instead of ideal gas pressure maintained by energy input
from fusion reactions.  This mass---the lowest mass at which a star
can fuse hydrogen---is now called the ``hydrogen burning mass limit''
(hereafter abbreviated as HBML).  Modern calculations of the HBML
place it, for objects of solar metallicity, between 0.080 $M_\odot$
and 0.070 $M_\odot$ (84 to 73 $M_J$, where $M_J$ is the mass of
Jupiter; Burrows and Liebert 1993, Baraffe et al.\ 1995).

\subsection{B. Definition of ``Brown Dwarf'' and of ``Planet''}

The canonical definition of a brown dwarf is a compact object, which
has a core temperature insufficient to support sustained nuclear
fusion reactions.  As shown above, this temperature requirement
translates directly into a mass requirement: a brown dwarf's mass must
be below the HBML.

This definition does not distinguish planets from brown dwarfs,
however.  Consensus in the literature on this issue suggests that
planets and brown dwarfs be distinguished by their formation
processes.  Planets form in circumstellar disks while brown dwarfs
form out of interstellar gas cloud collapse.  This distinction is
problematic because there is no simple observable of the birth
process.  In the case of brown dwarf companions of stars, one might
expect the orbit to be rather eccentric about the central star, while
the orbit of a planet might be roughly circular (Black 1997).
However, as reviewed by Lissauer et al. (this volume), planets formed
in circumstellar disks can be in highly eccentric orbits.

We propose a new definition of planets as objects for which no nuclear
fusion of any kind takes place during the entire history of the
object.  (As far as we know, the only other attempt to define the term
``planet'' in the literature is that of Basri and Marcy (1997), which
suggested a scheme similar to the one presented here.  Though Burrows
et al.\ (1997) use this same classification scheme, they present it as
a purely semantic definition only for the purposes of their paper and
do not advocate its replacement of the standard formation-motivated
definition.  We do.)  According to Burrows et al.\ (1997), objects (at
solar metallicity) with masses between 0.08 $M_{\odot}$ and 0.013
$M_{\odot}$ fuse deuterium when they are young.  Deuterium undergoes
fusion reactions at lower temperatures than hydrogen, primarily
because the reaction D($p,\gamma$)$^3$He is extremely rapid, being
driven by the electromagnetic force.  In contrast, the pp chain,
driven by the weak nuclear force, is much slower and therefore less
efficient, requiring higher temperatures.  A plot of the luminosity
evolution of objects between 0.2 $M_{\odot}$ and 0.0003 $M_{\odot}$
(Fig.\ 2) illuminates this issue.  The very highest mass objects,
stars, start out bright but eventually reach an equilibrium luminosity
at the right side of the plot.  The lower curves, for brown dwarfs and
planets, continue to drop in luminosity past 10$^{10}$ yr.  The first
bump in the upper curves of Fig.\ 2 indicates the age at which
deuterium fusion ends, having completely depleted the deuterium fuel.
However, for masses below 0.013 $M_{\odot}$, the curves are devoid of
this bump because the objects are not even capable of fusing
deuterium.  These objects are planets.  With these precise
definitions, planets, brown dwarfs and stars occupy a hierarchy based
on their internal physics.  Stars fuse hydrogen in equilibrium, brown
dwarfs do not fuse hydrogen in equilibrium but do fuse deuterium for
some portion of their evolution and planets never fuse anything.
Table I provides a summary of this hierarchy.

\smallskip

\centerline{\psfig{file=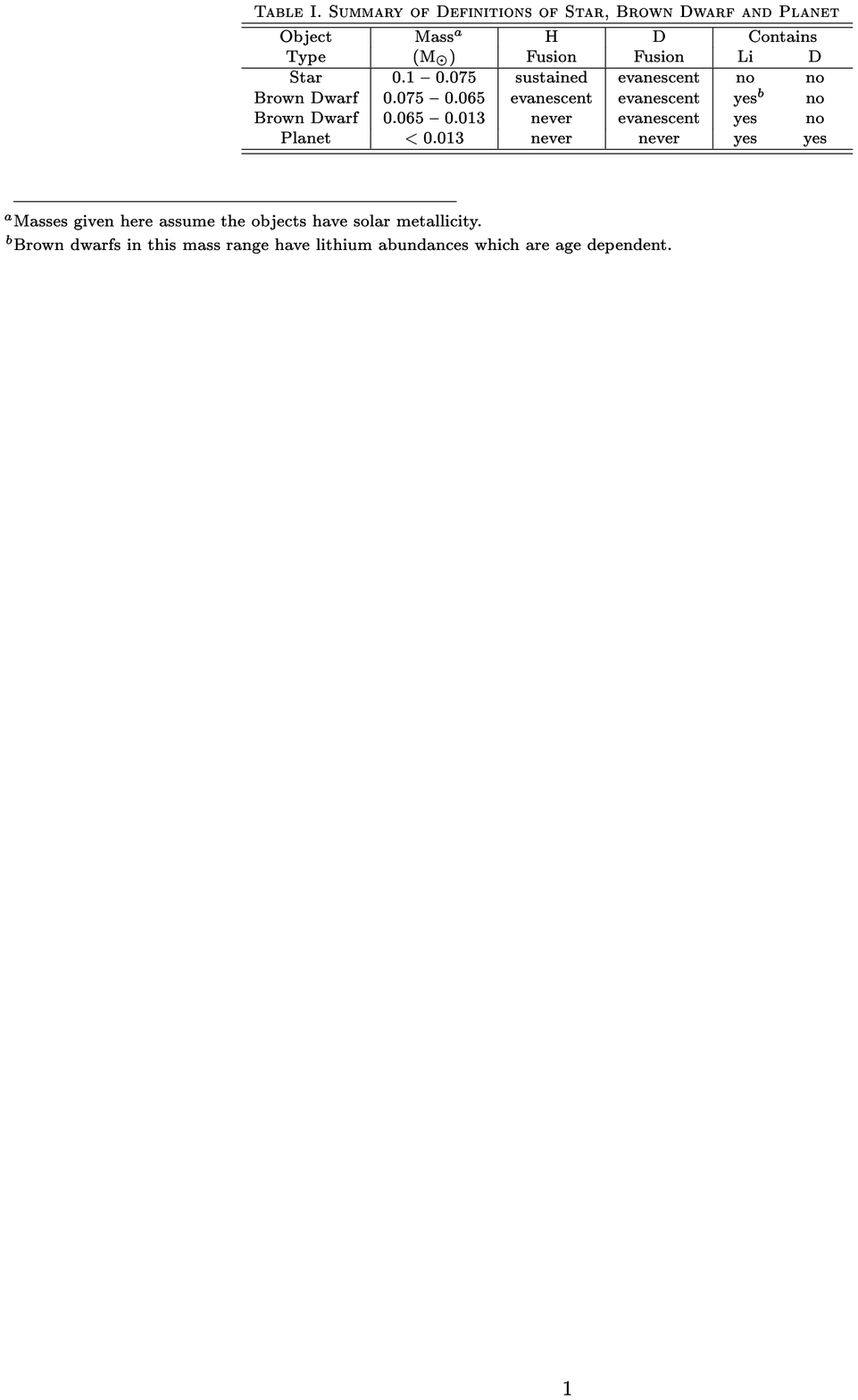,angle=0,width=4.3in}}

\smallskip

\centerline{\psfig{file=burr7.ps,angle=270,width=4in}}

\noindent
{\caption{Fig.\ 2 \capskip Cooling 
curves for objects with masses between 0.2 $M_\odot$ and 0.0003
$M_\odot$ from Burrows et al.\ (1997).  Stars, objects above 0.075
$M_\odot$ eventually reach a constant luminosity, where as brown
dwarfs and planets continue to cool throughout their existences.  The
first knee on the higher mass curves indicates the time when deuterium
fusion ends and ceases to be a source of energy.  The second knee
indicates where grain formation begins.  This results in a sudden
cooling of the objects.  Burrows et al.\ have called objects below 13
$M_J$ planets only because they never engage in deuterium fusion (see
\S I).  It is important to note that the minimum luminosity for a star
is 10$^{-4}$ L$_\odot$ which corresponds to $T_{\rm eff} = 1800$ K.
Courtesy A. Burrows.}}

\smallskip

Another justification of these definitions comes from the untested
theoretical notion that deuterium fusion and convection, which may
lead to magnetic fields and thus mass outflows, might halt the
accretion process as these low-mass objects form (Shu et al.\/ 1987).
This process would then place the lower limit to the mass of an object
formed in isolation (not in a circumstellar disk) at 0.013 $M_\odot$,
because a lower mass object could continue to accrete mass until it
exceeded this limit when deuterium burning, and consequently mass
outflow, would ensue.  Thus, planets could not form the way stars and
brown dwarfs do.  However, if confirmed, this would be an outcome of
the definition, not the overriding principle.

\subsection{C. Observational Identification of Brown Dwarfs}

There are three principal methods for confirming that a candidate
brown dwarf is, in fact, substellar.\footnote{$^{\dag}$}{In our
discussion of the observations of brown dwarfs we discuss only
directly detected objects.  For this reason we do not include
substellar objects discovered in radial velocity studies.  (See Marcy
et al.\ this volume.)}

\subsection{1. $L \propto T_{\rm eff}^4$}

The definitions presented above have direct consequences for the
observations of brown dwarfs, stars and planets.  The most obvious
distinction between stars and brown dwarfs is illustrated in the
cooling curves mentioned above (Fig.\ 2).  An object of solar
metallicity below 10$^{-4}$ $L_{\odot}$ cannot be a star, regardless
of its age.  Intrinsic luminosity, $L$, is an observable only for
objects with known distances.  A good, but less sensitive and yet more
practical, surrogate for luminosity is effective temperature, $T_{\rm
eff} = (L/4\pi R^2\sigma)^{1/4}$, where $\sigma$ is the
Stefan-Boltzmann constant, because, as we demonstrated above, $R$ is
essentially constant for objects below the HBML (except very low-mass
planets, where only the Coulomb force is important; see Fig.\ 1).
Spectral synthesis models are complete enough at this point that
comparison of spectra with the models constrains $T_{\rm eff}$ to
better than 10\%\ in most cases.  A luminosity of 10$^{-4} L_{\odot}$
corresponds to $T_{\rm eff} = 1800$ K.  The cooling curves show that a
0.013 $M_\odot$ brown dwarf reaches this temperature at an age of
approximately 100 Myr.  Therefore, this technique for identifying
brown dwarfs only works for relatively old and cool objects.

\subsection{2. Lithium}

Distinguishing young, hot brown dwarfs and planets from stars is
easiest with the ``lithium test.''

The lithium test as proposed by Rebolo et al.\ (1992) relies on the
fact that objects without hydrogen fusion retain their initial lithium
abundances forever.  This is a direct result of one of the nuclear
fusion reactions: Li$^7$($p,\alpha$)He$^4$.  This reaction effects the
complete destruction of lithium in the cores of very low-mass stars in
50 Myr and in brown dwarfs with masses between 0.08 and 0.065
$M_{\odot}$, which have short lived hydrogen fusion reactions, in 50
to 250 Myr (D'Antona and Mazzitelli 1994, Bildsten et al.\ 1997).
Below 0.065 $M_\odot$ brown dwarfs retain their initial lithium
abundances forever because they never host any hydrogen fusion
reactions.

Theoretical models show that brown dwarfs and very low-mass stars are
fully convective.  Thus, the elemental abundances in the core where
the putative fusion reactions happen are reflected on the convective
timescale in their observable atmospheres.  The convective timescale
for a brown dwarf is on the order of decades but scales proportional
to $L^{1/3}$.  In contrast the evolutionary timescale is 6 to 8 orders
of magnitude larger (Burrows and Liebert 1993, Bildsten et al.\ 1997),
so core abundances can be assumed identical to atmospheric abundances.

Fig.\ 3 (from Rebolo et al.\ 1996) shows lithium abundance
measurements as a function of $T_{\rm eff}$ for objects in the
Pleiades.  G and K stars have cosmic lithium abundances, but once
$T_{\rm eff}$ reaches the M dwarf regime, the lithium abundance
plummets for the reasons explained above.  Below 3000 K, young brown
dwarfs, such as PPL 15, Teide 1 and Calar 3 (described below), have
measurable lithium abundances.

\medskip

\centerline{\psfig{file=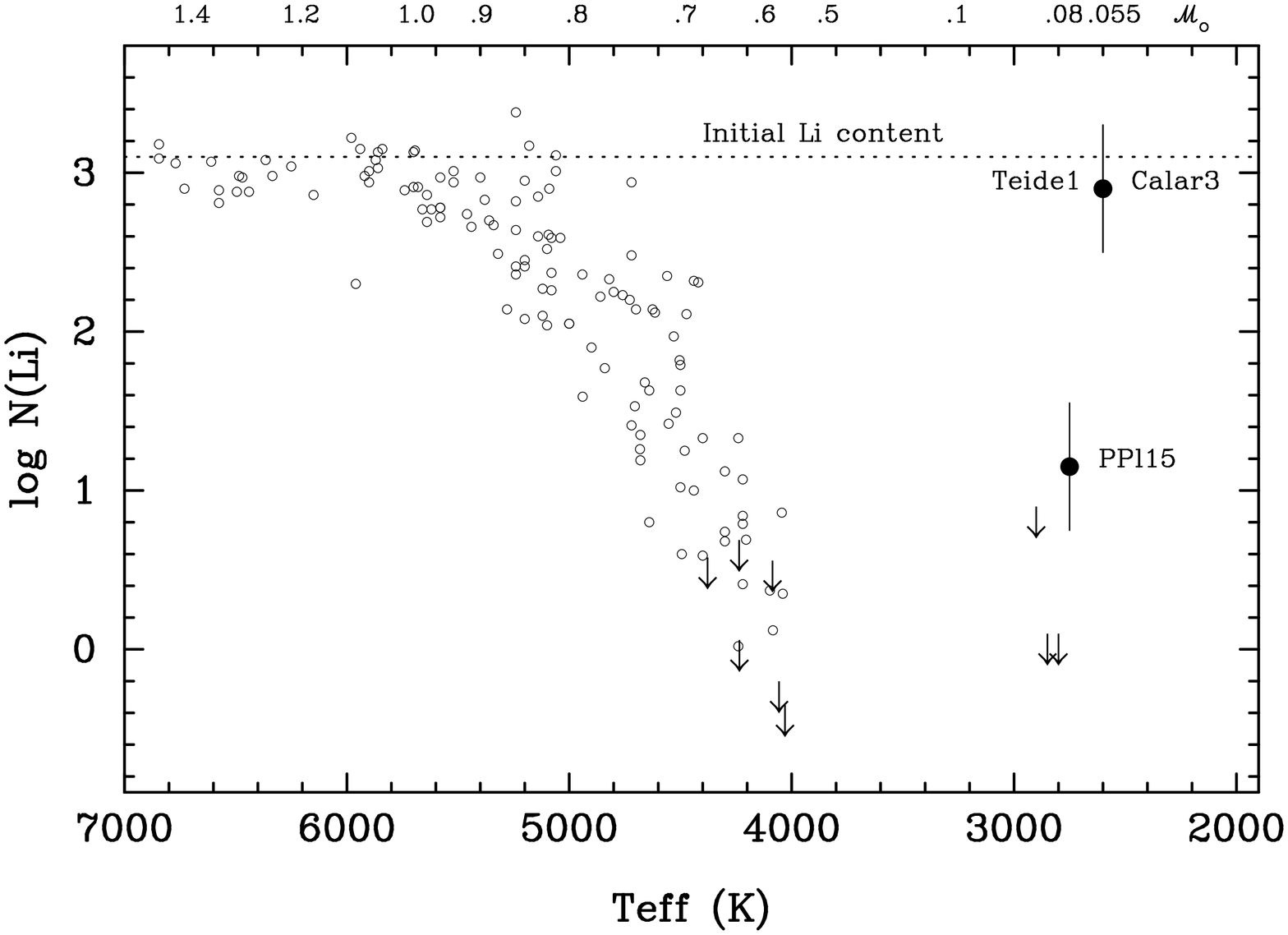,angle=270,width=4in}}

\noindent
{\caption{Fig.\ 3.\capskip Lithium abundance versus $T_{\rm eff}$ for
stars and brown dwarfs in the Pleiades.  This plot shows how the
presence of lithium in a low-temperature object can be used to
establish its classification as a brown dwarf.  The G and K dwarfs
have lithium, but M dwarfs do not, because they are fully convective
and the hydrogen fusion reactions destroy lithium.  Brown dwarfs
contain lithium because they do not support fusion reactions.  PPL 15,
Teide 1 and Calar 3, members of the Pleiades star cluster, are the
first brown dwarfs confirmed in this manner.  Courtesy of
M. Zapatero-Osorio, from Rebolo et al.\ 1996.}}

\smallskip

An interesting outcome of the lithium test is that one can accurately
determine the age of an open cluster by finding the ``lithium
depletion boundary,'' which is an imaginary line that separates faint
objects without lithium from slightly fainter objects with lithium.
This is demonstrated in Fig.\ 4 for the Pleiades (Stauffer et al.\
1998).  After 250 Myr, this boundary remains indefinitely at 0.065
$M_{\odot}$.  In older clusters, the brightest objects with lithium
will have a mass of 0.065 $M_{\odot}$, which can be used with the
measured luminosity to place the object in a well-constrained part of
Fig.\ 2.  Objects above the HBML deplete their lithium within 100 Myr,
so as long as the cluster being studied is older than 100 Myr, all of
the cool objects in the cluster which show lithium absorption must be
brown dwarfs.

Fig.\ 4 demonstrates the application of this technique to determine
the age of the Pleiades.  The lithium depletion boundary is indicated
by a line perpendicular to the zero-age main sequence and is defined
by extensive spectroscopic observations of all the Pleiads near the
line.  (See \S II.B for more discussion.)

\smallskip

\centerline{\psfig{file=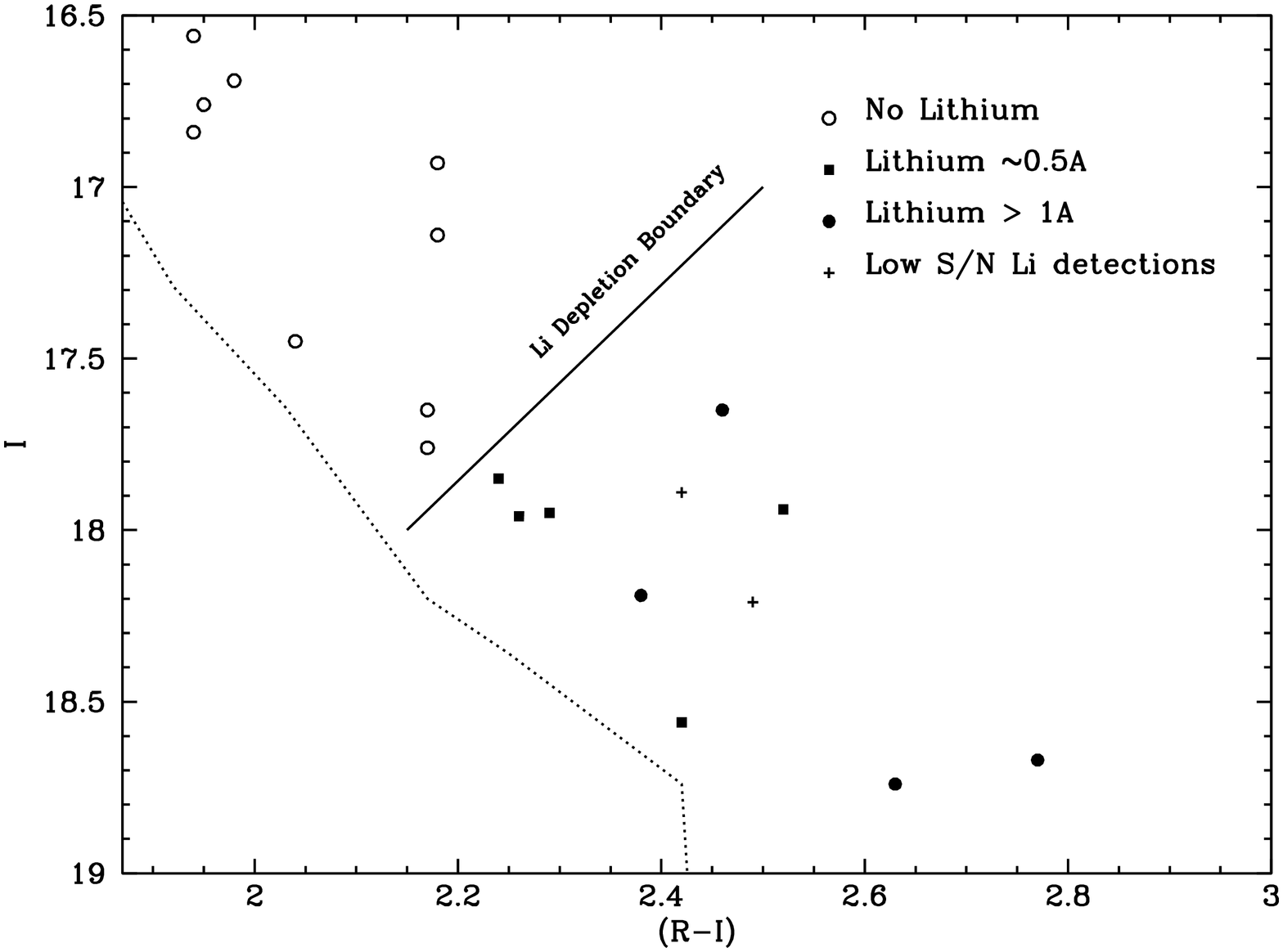,angle=270,width=4in}}

\noindent
{\caption{Fig.\ 4.\capskip Color-magnitude diagram for Pleiades very 
low-mass star and brown dwarf members with available spectra capable
of detecting the lithium 6708 \AA\ doublet at equivalent widths
greater than 0.5 \AA.  The dotted line is an empirical main sequence
at Pleiades distance.  The location of the ``lithium depletion
boundary'' is indicated by the solid line and is used to determine a
precise age for the cluster of 125 $\pm$ 8 Myr (Stauffer et al.\
1998).}}

\smallskip

\subsection{3. Molecules}

Once a brown dwarf cools to $T_{\rm eff} = 1500$ K, lithium begins to
form molecules and the Li I spectral signature weakens (Pavlenko 1998,
Burrows and Sharp 1998).  Fortunately a new diagnostic becomes
available.  Below 1500 K, chemical equilibrium between CO and CH$_4$
strongly favors CH$_4$ (Tsuji 1964, Fegley and Lodders 1994, Burrows
and Sharp 1998).  CH$_4$ has a number of extremely strong absorption
features in the range of 1 to 5 $\mu$m.  As a result, the
spectroscopic detection of methane means that the effective
temperature of the object must be below 1500 K, requiring that it be
less massive than the HBML (\S 1.C.1).  Ammonia forms at slightly
lower temperatures than 1000 K and a host of more exotic species appear
at even cooler temperatures.  See Burrows et al. (this volume) for a
more complete discussion of this progression.  These spectral
signatures allow observers to classify brown dwarfs by $T_{\rm eff}$.
In \S II and III we deal with this subject in greater depth.

\subsection{4. Deuterium}

Distinguishing brown dwarfs from planets, as defined here, involves a
search for deuterium.  Several of the planets in the solar system have
measured deuterium abundances.  (See, for example, Krasnopolsky et
al.\ 1997.)  In brown dwarfs, deuterium should be depleted to
unmeasurable quantities, even in their atmospheres because convection
causes a complete reflection of the core abundances in the atmosphere,
as we reasoned in \S I.C.2.  Spectroscopic signatures of deuterium
include absorption lines of HDO with numerous features between 1.2 and
2.1 $\mu$m (Toth 1997) and possibly CH$_3$D with strong features at
3.7 and 4.4 $\mu$m (Noll 1993, Krasnopolsky et al.\ 1997) in the 1 to
8 $\mu$m region.  This proposed classification scheme is only hampered
by the current state of technology, in that spectra of the known
extrasolar ``planets'' indirectly detected through radial velocity
studies (Marcy et al.\ this volume) cannot be obtained yet.

\subsection{Caveat: Dust}

As brown dwarfs cool, theory predicts that dust will form in the
atmosphere.  Even some cool main-sequence stars seem to contain dust
(Jones and Tsuji 1997).  Dust formation occurs at the ridge in Fig.\ 2
at approximately 10$^{-4} L_{\odot}$ (1800 K).  As the dust forms, the
luminosity drops more precipitously.  Below this ridge, a progression
of species with important features in the near IR appears.  These
species will have two effects: (1) weakening of the molecular
absorption features and (2) reorganization of the broad-band spectral
energy density toward a black body spectrum.

\subsection{D. Observational History}

Ever since Kumar's pioneering work, astronomers have searched for
brown dwarfs primarily because they were regarded as ``terra
incognito.''

Some of the initial discussions of brown dwarfs suggested that they
could be the ``missing'' matter implied by the dynamics of the galaxy.
For example, simple extension of the Salpeter initial mass function
(IMF), in which $dN/dM \propto M^{-2.35}$, to brown dwarf masses
suggests that brown dwarfs ought to outnumber stars by two or three
orders of magnitude.  Whether an appreciable percentage of the dark
matter is brown dwarfs or planets is still the subject of some debate.
Most researchers agree that based on the microlensing experiments of
MACHO (Alcock et al.\ 1998), most of the dark matter is not
made of brown dwarfs.  However, by adopting unusual parameters, it is
still possible to construct galactic models consistent with the MACHO
results and with more than 50\%\ of the dynamical mass in brown dwarfs
(Kerins and Wyn Evans 1998).

From 1984 through 1994 approximately 170 refereed papers were written
on brown dwarfs.  By the end of 1996 that number doubled, and based on
current publication rates, 1998 alone will see approximately 170 more
papers submitted.  This sudden explosion in observational and
theoretical results was due to the discovery in late 1995 of Gliese
229B, the first cool brown dwarf detected (described in \S III), and
the confirmation of lithium in the brown dwarf candidate PPL 15 in
early 1996 (and in Teide 1 and Calar 3 slightly later).  The sustained
publication rate is largely due to new large scale surveys which are
now turning up brown dwarfs by the dozen.  These dramatic successes,
however, were preceded by several decades of unsuccessful searches and
two conferences whose proceedings are punctuated with the wreckage of
disproven brown dwarf candidates and steady improvements in
theoretical work.

In 1985, when the first conference solely devoted to brown dwarfs was
held at the George Mason University (Kafatos et al.\ 1986), a new
breed of infrared and optical detectors had enabled the first searches
designed to detect brown dwarfs directly.  No brown dwarfs were found,
however, and in retrospect this is because of a lack in sensitivity.
The basic strategies behind these early searches are imitated to this
day.  One can look for brown dwarfs in isolation or as companions of
nearby stars.  Isolated brown dwarfs can be found in all-sky surveys
or smaller surveys of star clusters.  Companion searches employ
techniques to prevent the bright nearby star from washing out the
faint companion.

In 1987 the results of the Infrared Astronomical Satellite (IRAS),
launched in 1984 to survey the entire sky at far infrared wavelengths,
were presented.  One of the principal goals of this satellite mission
was to find brown dwarfs.  None was detected (Beichman 1987).

The first direct searches for brown dwarf companions of nearby stars
were coincident with the development of sensitive electronic infrared
photodetectors.  Probst (1983) used a single pixel device on NASA's
Infrared Telescope Facility (IRTF) to search for infrared excess
around nearby white dwarf stars.  The search was sensitive to
companions within 15 arcsec of the stars.  Probst targeted white dwarf
stars because they are intrinsically fainter than main sequence stars,
so that any infrared excess would be easier to detect.  No brown
dwarfs were detected.

Using the same single pixel detector, McCarthy et al.\ (1985) applied
the technique of speckle interferometry.  This technique uses rapid
exposures to compensate for the blurring effects that the turbulent
atmosphere has on astronomical images.  In principle this permits the
detection of a companion fainter and closer to the primary star than
is possible in a standard direct image.  McCarthy et al.\ (1985)
reported a faint companion of the red dwarf, vB 8.  Their results
suggested that this object had a luminosity of about $10^{-5}
L_\odot$.  However, the putative companion was never detected again
and remains an irreproducible result.

Forrest et al.\ (1988), using a 32$\times$32 pixel InSb array on the
IRTF, found several stellar companions of red dwarfs in the solar
neighborhood, but still uncovered nothing faint enough to be
considered a brown dwarf.

In 1988, Becklin and Zuckerman, who extended the work of Probst (1983)
to survey nearby white dwarfs for faint companions, found the object
known as GD 165B.  GD 165B has a temperature of 1800 K (Jones et al.\
1994), and until 1995 was the best candidate brown dwarf known.
Subsequent spectroscopy by Kirkpatrick et al.\ (1998) has demonstrated
that GD 165B has no lithium and is therefore either a star right at
the HBML or a high mass brown dwarf a few Gyr old.

Henry and McCarthy (1990) conducted a search for infrared companions
of the stars within 8 pc of the sun using the speckle interferometry
technique and an array of pixels.  Although they found no brown
dwarfs, their sensitivity was somewhat limited, reaching a maximum of
7.5 magnitudes of difference between the central star and the faintest
object detectable (Henry et al.\ 1992).  For reference, Gliese 229AB
(a red dwarf-cool brown dwarf system) has a contrast of over 10
magnitudes in the near infrared (Matthews et al.\ 1996; see \S III).

Substantial gains in detecting stars of lower and lower mass had been
made by the next conference on brown dwarfs, held in Garching (see
Tinney 1995 for proceedings).  However, still no definitive brown
dwarfs were presented.  Determinations of the mass function presented
at this meeting suggested that brown dwarfs are extremely rare, but,
in retrospect, the surveys used were still not sensitive enough
(D'Antona 1995).

During the course of a coronagraphic survey of nearby stars, Nakajima
et al.\ (1995) reported the discovery of an object with the same
proper motion as Gliese 229.  At a distance of 5.7 pc from the Sun,
the inferred intrinsic luminosity of the companion, Gliese 229B, is
6.4 $\times 10^{-6} L_{\odot}$ (Matthews et al.\ 1996).  Spectroscopy
of this object revealed deep methane features and implied a
temperature below 1200 K (Oppenheimer et al.\ 1995, 1998).  Although
many astronomers had become inured to brown dwarf announcements which
were retracted months later, the spectrum of Gliese 229B is
sufficiently distinctive that, when shown briefly at the 9th ``Cool
Stars, Stellar Systems and the Sun'' conference in Florence, Italy
(1995), it was unanimously taken as proof that the object was indeed a
brown dwarf.

Parallel to the searches for companion brown dwarfs, several search-es
for isolated brown dwarfs were conducted between 1989 and 1995.  The
first CCD deep imaging surveys of the Pleiades believed to have
reached below the HBML were those of Jameson and Skillen (1989,
hereafter JS89) and Stauffer et al.\ (1989, hereafter S89).  These
initial surveys only covered very small portions of the cluster: JS89
imaged just 225 square arcmin and identified seven objects as
likely Pleiades brown dwarfs, while S89 surveyed 1000 square
arcmin and identified just four brown dwarf candidates.
Subsequent analysis indicated that due to an error in the photometric
calibration, only one of the JS89 objects was faint enough to be a
possible Pleiades brown dwarf (Stringfellow 1991; Stauffer et al.\
1994).

After the lithium test was proposed in 1992, several attempts to apply
it to brown dwarf candidates in the Pleiades revealed no lithium
(Mart\'{\i}n et al.\ 1994, Marcy et al.\ 1994).  This was largely
because the Pleiades is older than these studies presumed, so they
selected candidates that were not faint enough.  By obtaining a
spectrum of a fainter object, PPL 15 (a Pleiades brown dwarf candidate
identified by Stauffer et al.\ 1994), Basri et al.\ (1996, hereafter
BMG) made the first detection of lithium in a brown dwarf candidate.
Rebolo et al.\ (1996, hereafter R96) soon after detected lithium in
two other Pleiades brown dwarf candidates, Teide 1 and Calar 3.

By the beginning of 1996 the first brown dwarfs had been found, and in
March 1997 a conference was held in Tenerife, Spain (Rebolo et al.\
1998), where the wealth of positive observational results was a direct
testament to the sudden change in the field.  We separate our
discussion of successful observations of brown dwarfs into two
sections, one on isolated brown dwarfs (\S II) and the other on brown
dwarf companions of nearby stars (\S III).  We also recommend the
useful reviews by Kulkarni (1998), Allard et al.\ (1997), Hodgkin and
Jameson (1997).

\mainsection{{I}{I}.  ISOLATED BROWN DWARFS}
\backup

The principal reason for studying isolated brown dwarfs is to acquire
a complete census of objects with masses below the HBML (i. e. to
measure the mass function).  The relative number of brown dwarfs of a
given mass compared to the number of higher mass objects has important
implications for star formation theories.  Indeed, the eventual mass
of an object formed out of an interstellar cloud fragment would seem
to be entirely independent of the HBML, so that objects with masses
well below the HBML ought to form out of interstellar cloud
fragmentation (Burkert and Bodenheimer 1993; Shu et al.\ 1987).
Observations of star formation regions in which very low mass clumps
of gas exist certainly suggest that brown dwarfs can form out of this
process (Pound and Blitz 1995).  Measuring the mass function would
determine whether there is a lower limit to the mass of an object
formed like a star and not in a circumstellar disk.

Because brown dwarfs cool, and a given brown dwarf can have a huge
range of luminosities over its lifetime, making a complete census of
them is greatly simplified by examining a sample of the same age.  In
such a sample, mass will be solely a function of luminosity, which is
directly observable as discussed in \S I.C.1.  By far the best means
to find a population of brown dwarfs with the same age is to identify
low luminosity members of well-studied open clusters where, in
principle, the age, distance and metallicity should be known
accurately.

Surveys for field brown dwarfs, outside stellar associations, require
immense sky coverage because of the intrinsic faintness of the brown
dwarfs, which effectively limits the volume of space that the surveys
probe.  For example a 1000 K brown dwarf is detectable by the new near
infrared all sky surveys (2MASS and DENIS; see below) out to a
distance of only about 6 pc.

Critical to both types of surveys is the certification that a given
object is a brown dwarf.  The principal method for this is the lithium
test.  However, as the surveys probe fainter and fainter limits,
certification through the molecular features described in \S I.C.3
will become equally important.

\subsection{A.  Brown Dwarfs in Open Clusters and Star Forming Regions}
Brown dwarf candidates are identified in photometric surveys based on
their lying above the zero age main sequence (Fig.\ 4).  To confirm
that they are in fact cluster members (and not reddened, background
stars) requires accurate proper motion measurements or accurate radial
velocity measurements.  However, to date at most one of the likely
cluster brown dwarfs has a sufficiently accurate proper motion (Rebolo
et al.\ 1995).  The lithium test has therefore provided the primary
means to confirm the substellar nature of the young brown dwarf
candidates.

\centerline{\psfig{file=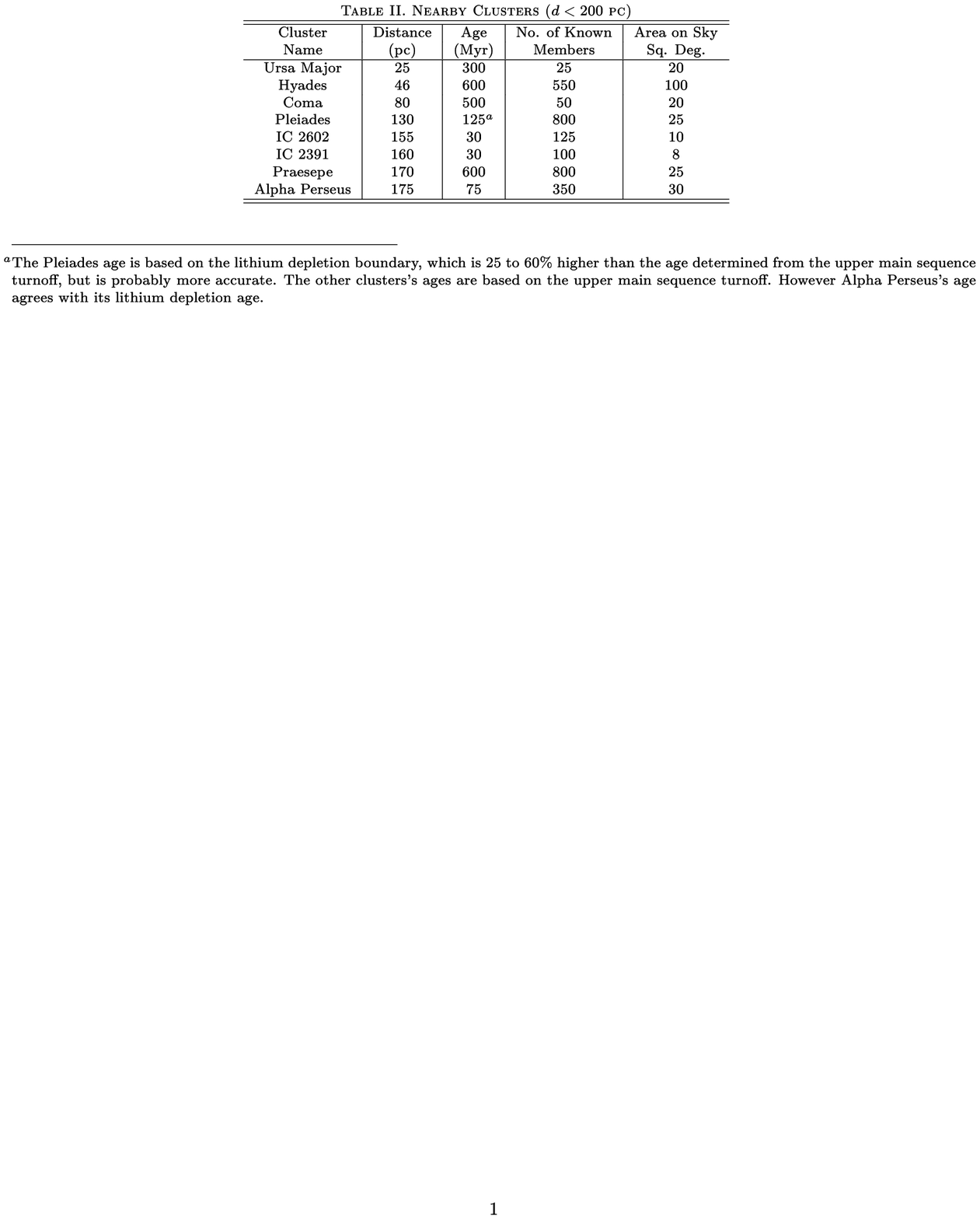,angle=0,width=4.3in}}

\subsection{1.  The Substellar Mass Population of the Pleiades}

The Pleiades is the richest, nearby open cluster.  (See Table II.)
For a nominal age of 100 Myr (Meynet et al.\ 1993), objects at the
HBML are predicted to have effective temperatures of about 2500 K,
corresponding to spectral class M6 V on the main sequence.  Because of
the cluster's proximity to the Sun (Table II), these brown dwarfs
should be detectable with modern optical CCDs.  The
cluster half mass radius is $\sim$2 pc, and the tidal radius is about
16 pc (Raboud and Mermilliod 1998; Pinsonneault et al.\ 1998).  The
areas on the sky corresponding to circles with these radii are 2.5 and
150 square degrees, respectively.  This is important because it
indicates that it is necessary to search a large area to sample a
significant portion of the cluster.  Because the Hyades is three times
closer, it is spread over a much larger area on the sky than the
Pleiades.  For these reasons, the Pleiades has been the principal
hunting ground for isolated brown dwarfs.
 
Since 1989, at least 10 deep imaging surveys of the Pleiades other
than those described in \S I.D have been conducted.  A summary is
provided in Bouvier et al.\ (1998).  By using redder filters and more
sensitive, larger format CCDs, these surveys have been able to reach
lower inferred mass limits and cover larger portions of the cluster.
A conservative assessment of the current surveys suggests that at
least 40 substellar members of the Pleiades have now been identified
(Fig.\ 4).

Using the ``lithium depletion boundary'' described above, BMG and R96
estimated the age of the Pleiades at about 120 Myr with PPL 15 and HHJ
3 (Hambly et al.\ 1993), the faintest Pleiad without lithium, defining
the boundary.  However, Basri and Mart\'{\i}n (1998) subsequently
discovered that PPL 15's luminosity was over-estimated because it is an
approximately equal mass, short-period binary, with each component
being about 0.7 mag fainter than the composite and having a mass of
approximately 0.06 $M_{\odot}$.  PPL 15 is the first brown dwarf binary
system found.

Spectra of 10 additional Pleiades brown dwarf candidates have recently
allowed Stauffer et al.\ (1998) to define the lithium depletion
boundary in the Pleiades to $\pm$0.1 mag and thus to derive an age for
the cluster of $\tau$ $\sim$ 125 $\pm$ 8 Myr (Fig.\ 4).  By
coincidence, at this age the lithium depletion boundary corresponds to
0.075 $\pm$ 0.005 $M_\odot$, and therefore all Pleiades members
fainter than the lithium depletion boundary are brown dwarfs (Ventura
et al.\ 1998, Chabrier and Baraffe 1997).  The faintest Pleiades
candidates identified to date have masses on the order of 0.035
$M_\odot$ (Mart\'{\i}n et al.\ 1998).
 
Zapatero-Osorio et al.\ (1997) and Bouvier et al.\ (1998) have used
their surveys to estimate the Pleiades mass function in the substellar
regime.  Both groups obtain slightly rising mass functions for the
range 0.045 $\leq$ M $\leq$ 0.2 $M_\odot$, with dN/dM $\propto$\
M$^{-1.0}$\ and M$^{-0.7}$, respectively.  However, the relatively
small fraction of the cluster that has been surveyed to date make
these estimates fairly uncertain.

\subsection{2.  Brown Dwarfs in Other Open Clusters}

Basri and Mart\'{\i}n (1999) have reported a lithium detection for
the faintest known member of the $\alpha$ Persei open cluster (and
non-detection of lithium in one or two brighter, probable members),
thus allowing them to place the age of the cluster between 60 and 85
Myr (Table II).  

Two deep imaging surveys of Praesepe (see Table II) have been conducted
(Pinfield et al.  1997; Magazz\`u et al.\ 1998).  Magazz\`u et al.\
report one object in their survey with I $\sim$ 21 and a spectral type
of about M8.5, which would indicate a mass near the substellar limit
if the object is indeed a Praesepe member and if the cluster age is as
expected ($\sim$600 Myr).

The deepest survey of the Hyades to date is that provided by Leggett
and Hawkins (1988) and Leggett et al.\  (1994).  The faintest objects
in this survey may also be approximately at the substellar mass
boundary; however, no spectra for the faintest candidates have yet
been reported.

\subsection{3.  Brown Dwarfs in Star Forming Regions}

Brown dwarfs in star-forming regions (age $<$ 1 Myr) will be much more
intrinsically luminous than those in the open clusters discussed above
and should be easier to discover.  However, it is in fact more
difficult to ``certify" that any given object is substellar in a
star-forming region than in an open cluster.  First, the ``lithium
test" is of limited value at this age because all low mass stars
should still have their original lithium abundance.  However, as Basri
(1998) points out, if a candidate object lacks lithium it can be
discarded as a member of the star forming region.  Second, the
theoretical isochrones for young, low mass objects are quite
uncertain.  Thus, determining the ages of these objects is difficult.
Even if the age can be determined, the intrinsic luminosity of a
candidate object is difficult to measure because of uncertainties in
extinction parameters for these star forming regions.

However, based on the existing models and using the Pleiades as a
reference, Basri (1998) has argued that any object with spectral type
M7 or later must be a brown dwarf if it contains lithium.  This is
because stars more massive than the HBML deplete lithium before they
can cool to the M7 effective temperature (i. e. before they reach the
zero-age main sequence).  Brown dwarfs on the other hand can cool to
the M7 spectral type when they are much younger than the stars and
still retain their lithium.
 
Luhman et al.\ (1997) have identified an apparent brown dwarf member of
the $\rho$\ Ophiuchus star-forming region based on its spectral type of
M8.5.  This object was originally thought to be a foreground star
(Rieke and Rieke 1990); however, the new data indicate that it is much
more likely to be a member of the cluster (in particular, it has very
strong \Ha\ emission and relatively low surface gravity).  Comeron et
al.\ (1998) have identified 3 other members of this region with
spectral classes $>$ M7, based on new data with the ISO satellite and
spectroscopy by Wilking et al.\ (1998).

Luhman et al.\ (1998) have also obtained spectra for a large number of
faint candidate members of the star forming region IC348, and have
identified three good brown dwarf candidates---two with spectral type
M7.5 and one with spectral type M8. 

\subsection{C.  Brown Dwarf Members of the Field Population}

The search for isolated brown dwarfs in the field has also seen
dramatic progress in the past 2 years.  The primary sources of the
newly discovered field brown dwarfs are the wide-field, near-IR
imaging surveys DENIS and 2MASS; however, a number of objects have
also been identified using other techniques.

\subsection{1.  Brown Dwarfs from DENIS and 2MASS}

DENIS (DEep Near-Infrared Survey) obtains simultaneous images at I, J
and K of the southern sky to limiting magnitudes of 18.5, 16 and 14.0
(3$\sigma$), respectively.  Based on the photometry of previously
identified very-low mass stars, the DENIS project uses a color
criterion of I$-$J $>$ 2.5 to select ``interesting" objects, with the
most interesting objects being those with colors like that of GD165B.
The DENIS search contains no color criteria to distinguish analogs of
Gliese 229B.

The DENIS team has reported about 5 objects with GD165B colors after
analyzing only 500 square degrees.  Optical spectra have been obtained
for three of those objects, with one of them showing a strong lithium
absorption feature (Delfosse et al.\ 1997; Tinney et al.  1997;
Mart\'{\i}n et al.\ 1997).  All of these objects have spectra in the
0.8 $\mu$m region similar to that of GD165B.  The lithium feature in
combination with the very late spectral type for DENIS-P
J1228.2$-$1547 indicate that this object is undoubtedly a brown dwarf.
No lithium has been detected in two of the objects with very late
spectral types.  They could be old substellar
objects in the mass range 0.065 to 0.075 $M_\odot$.

2MASS (Two-Micron All Sky Survey) obtains simultaneous images at J, H
and K of the entire sky to limiting magnitudes of 17, 16.5 and 15.5
(3$\sigma$), respectively.  Digitized scans of the E or N plates from
the Palomar Sky Survey are used to derive R or I magnitudes for
objects detected in the infrared.  2MASS uses color criteria of J$-$K $>$
1.3 and R$-$K $>$ 6, or J$-$K $<$ 0.4 to select brown dwarf candidates,
with the latter criterion being designed to find analogs of Gliese
229B (see \S III).  

More than a dozen candidates from 420 square degrees have had
spectroscopic follow-up observations (Kirkpatrick et al.\ 1998).  Six
of these objects have extremely late spectral type and show lithium
absorption, and thus are substellar; an approximately equal number of
objects are similarly late but do not show lithium.  None of the
objects observed spectroscopically show methane in their spectra, and
none have been found with colors similar to those of Gliese 229B.

Considering the small fraction of the sky analyzed so far, it appears
likely that 2MASS and DENIS will eventually provide a list of hundreds
of field brown dwarfs.  Due to the correlation of age, mass, effective
temperature and luminosity, it is inevitable that this sample of brown
dwarfs will favor relatively young objects with masses not far below
the HBML.  

The spectra of the coolest DENIS and 2MASS objects are sufficiently
different from previously known objects that a new spectral class must
be defined.  Kirkpatrick (1997) and Mart\'{\i}n et al.\ (1997) have
suggested use of the letter L for this class.  Kirkpatrick et al.\/
(1998) have begun to define this class through the weakening of TiO
bands to later types, the presence of resonance lines of alkali
metals---in particular potassium, rubidium and cesium, with these
lines becoming extremely strong at later types---the presence of other
molecular species such as CrH, FeH and VO, and the absence of methane.
One brown dwarf, Gliese 229B (\S III) does not fit in the L class
because it is several hundred degrees cooler than the coolest L dwarf.
Another spectral class must be created once Gliese 229B analogs are
discovered.

\subsection{2.  Other Field Brown Dwarfs}

Two field brown dwarfs have been identified from proper motion
surveys.  The first of these, Kelu-1, was identified as part of a
survey of 400 square degrees of the southern sky using deep Schmidt
plates (Ruiz et al.\ 1993).  Kelu-1 has a spectrum similar to GD165B's
and has lithium in absorption and \Ha\ in emission.  At K = 11.8,
Kelu-1 is comparatively very bright (and hence presumably quite
nearby), and so is a good target for detailed study.

The other field brown dwarf identified via proper motion is LP
944$-$20, originally catalogued as a high proper motion object by
Luyten and Kowal (1975).  It was rediscovered in a search for very red
objects by Irwin et al.\ (1991) and identified as a very late-type M
dwarf by Kirkpatrick et al. (1997).  Tinney (1998) subsequently showed
lithium was present with an equivalent width of about 0.5 \AA.  LP
944$-$20 has a measured parallax from which the intrinsic luminosity
can be derived.  Tinney used the inferred luminosity of 1.4 $\times$
10$^{-4}$ L$_{\odot}$ combined with an estimate of the lithium
abundance to derive a mass estimate of 0.06 $\pm$ 0.01 $M_\odot$ and
an age of about 500 Myr.

One other possible field brown dwarf has been identified from a deep
photographic RI imaging survey by Thackrah et al.\ (1997).  The
object, 296A, was selected because it was quite bright (I $\sim$14.5)
and reasonably red (R$-$I $\sim$\ 2.5).  Spectroscopy revealed a
spectral type of M6 and a lithium absorption equivalent width of about
0.5 \AA.  These features suggest that it could be a Pleiades age star
at the HBML.

\mainsection{{I}{I}{I}.  COMPANION BROWN DWARFS}
\backup

Searching for brown dwarf companions of nearby stars is attractive
mainly because it is the most effective way to identify the coolest
(lowest luminosity; see \S I.C.1) objects.  The principal difficulty
in this is the scattered light of the primary star.  There are several
techniques to circumvent this difficulty.  First, one can search at
the longer wavelengths, where the contrast between the brown dwarf and
the star is lowest.  Second, one can search for companions of white
dwarf stars where the contrast is very small because of the intrinsic
faintness of the star.  Third, the use of a coronagraph artificially
suppresses the starlight with a series of optical stops (Nakajima et
al.\ 1994).

\subsection{A.~~ Gliese 229B}

A large survey of nearby stars using a coronagraph (with a tip-tilt
image motion compensator) was carried out at the Palomar 60-inch
telescope.  This proved successful when Nakajima et al.\
(1995) showed that the star Gliese 229 has a companion with a
luminosity of less than $10^{-5} L_\odot$.  The discovery image of
Gliese 229B is shown in Fig.\ 5 with a subsequent image taken by the
Hubble Space Telescope (Golimowski et al.\ 1998).  The impact of this
discovery was far-reaching, not only did it validate the immense
effort of the astronomers who persisted in working on brown dwarfs
despite all the non-detections, but it also excited considerable
interest among planetary scientists, partly because of the fact that
Gliese 229B is orbiting a nearby star, but also because its spectrum
(Oppenheimer et al.\ 1995, 1998) looks remarkably like Jupiter's, with
major features due to water and methane.  Methane dissociates at
temperatures above between 1200 and 1500 K, a fact which further
implies the extremely low luminosity of Gliese 229B.

Matthews et al.\ (1996) made photometric measurements of Gliese 229B
from $r$ band at 0.7 $\mu$m through $N$ band at 12 $\mu$m.  These data
account for between 75 and 80\% of the bolometric luminosity of the
brown dwarf (depending on the model used for the unmeasured flux).
The observed luminosity is $(4.9 \pm 0.6) \times 10^{-6} L_\odot$,
implying a bolometric luminosity of $6.4 \times 10^{-6} L_\odot$ and
an effective temperature of 900 K assuming the brown dwarf radius is
$0.1 R_\odot$, as was argued in \S 1.A.

Allard et al.\ (1996) and Marley et al.\ (1996), using the photometry
of Matthews et al.\ (1996) and the spectrum from Oppenheimer et al.\
(1995) constrained the mass of Gliese 229B between 0.02 and 0.055
$M_\odot$ by fitting non-gray atmosphere models.  The mass is so
uncertain because (1) the age of Gliese 229B is unknown and
could be from 0.5 to 5 Gyr based on the spectrum of Gliese 229A and
(2) the gravity has not been accurately measured.

\smallskip
\centerline{\psfig{file=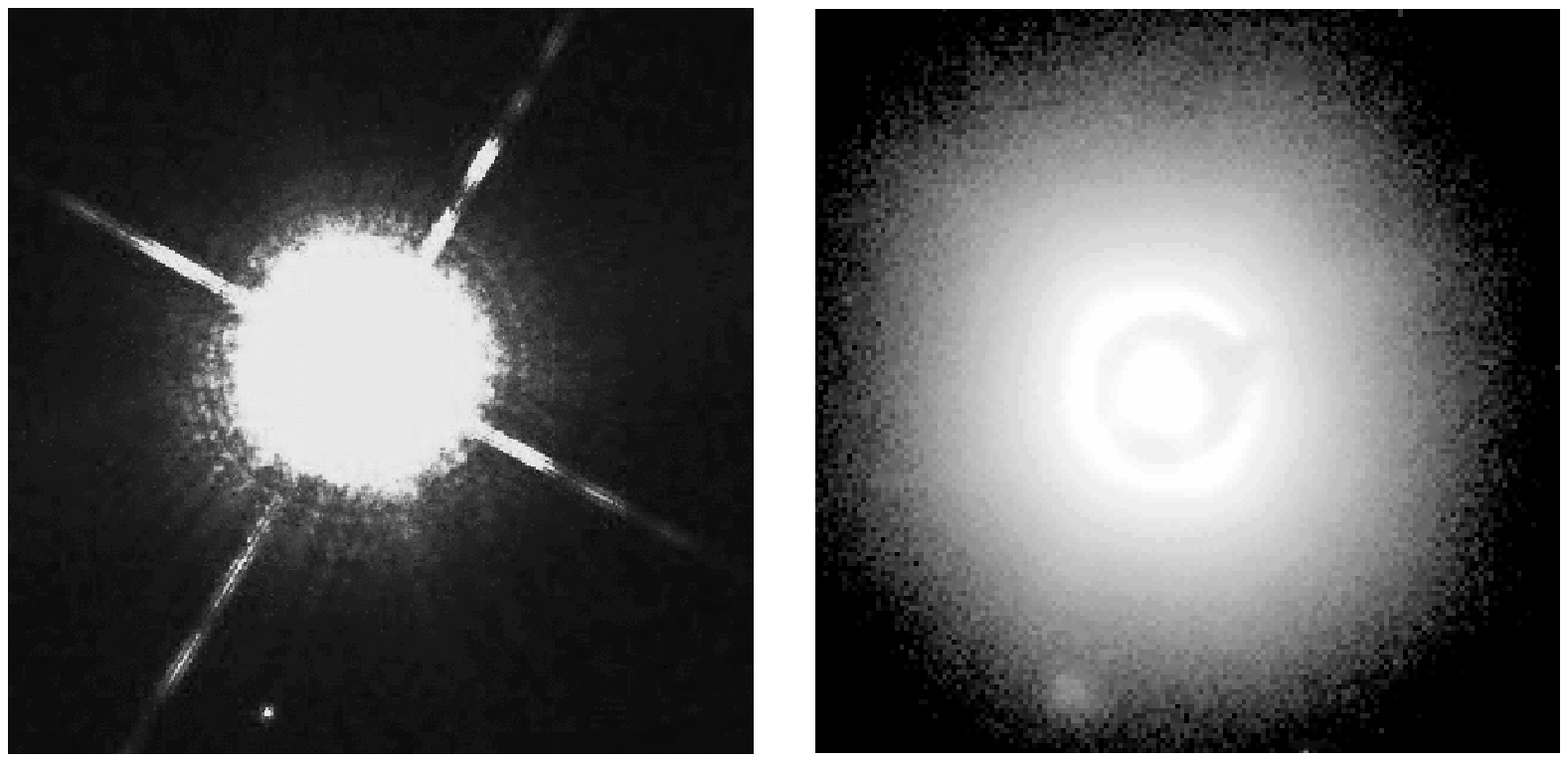,angle=270,width=4in}}

\noindent
{\caption{Fig.\ 5.\capskip Two images of the Gliese 229 system.  The
left panel shows a direct image from HST's WFPC2.  The brown dwarf is
at the bottom left of the image.  The right panel shows the discovery
image from the Palomar 60 inch telescope fitted with a coronagraph.
The coronagraphic stop is visible, obscuring most of the light from
the primary star.  The stop is 4 arcsec in diameter and is semi
transparent.  The brown dwarf is visible in the bottom left of the
image.  Both images are oriented with N up and E to the left and are
approximately 17 arcseconds on a side.}}

\smallskip

Geballe et al.\ (1996) obtained a high resolution spectrum of Gliese
229B in the 1 to 2.5 $\mu$m region and showed that there are hundreds
of very fine spectral features due to water molecules.  These may
be the most gravity sensitive features in the spectrum (Burrows et
al.\ 1997) and with higher resolution spectra the gravity of Gliese
229B may be constrained to within 10\%.

Another important conclusion of the Matthews et al.\ (1996) paper is
that the photometry indicates a complete lack of silicate dust in the
atmosphere of the brown dwarf.  Fig.\ 6 (from Matthews et al.\ 1996)
is a plot of the photometric measurements, along with a model spectrum
from Tsuji et al.\ (1996) and three black body curves assuming the
brown dwarf has a radius of $0.1 R_\odot$ and is at 5.7 pc.  The model
spectrum has no dust included in the calculation.  By adding even a
minute quantity of silicate dust, Tsuji et al.\ (1996) find that the
spectrum no longer fits the photometric data.  In contrast, the L
dwarfs of Kirkpatrick et al.\ (1998) and the spectrum of GD 165B
(Jones et al.\ 1998; Jones and Tsuji 1997) appear to be considerably
affected by the presence of dust.

Oppenheimer et al.\ (1998a) confirmed these conclusions with a high
signal-to-noise spectrum of Gliese 229B (shown in Fig.\ 7) in the near
infrared but also reported a smooth, almost featureless spectrum in
the optical (0.85 to 1.0 $\mu$m) region that could not be fitted by
any of the models.  An optical spectrum was also obtained from the
Hubble Space Telescope (HST) by Schultz et al.\ 1998.  There is
excellent overall agreement between the Keck and the HST spectra.
However the HST spectrum lacks the resolution and sensitivity to
reveal the fine features seen in the Keck spectrum (Fig.\ 7) and
discussed here.  

\smallskip
\centerline{\psfig{file=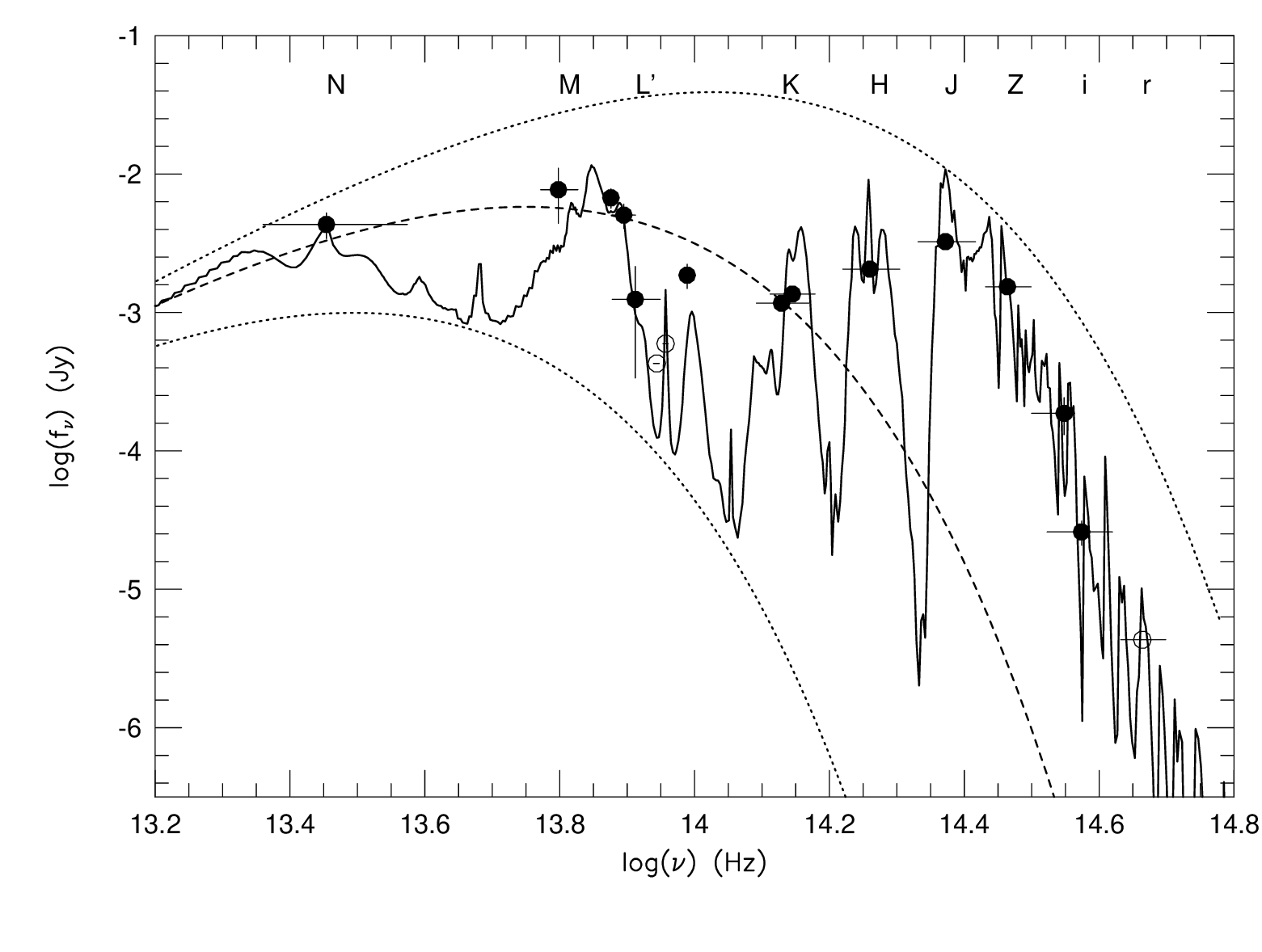,angle=0,width=4in}}

\noindent
{\caption{Fig.\ 6.\capskip Photometric measurements of Gliese 229B
(from Matthews et al.\ 1996).  The photometry shown here represents 75
to 80\% of the bolometric luminosity ($6.4 \times 10^{-6} L_\odot$) of
the brown dwarf and that the atmosphere is devoid of dust grains that
affect the infrared spectrum.  The solid line is a dust-free model
spectrum from Tsuji et al.\ (1996).  The long-dashed line indicates a
black body spectrum for $T_{\rm eff} = 900$ K, the estimated value for
Gliese 229B.  The dotted curves are black body spectra for $T_{\rm
eff} = 500$ K (bottom) and 1700 K (top).}}

\smallskip

The fact that the optical spectrum is smooth (Fig.\ 7) and that the
water band at 0.92 $\mu$m is shallower than expected (Allard et al.\
1997) indicates that there may be an additional source of continuum
opacity not previously discussed by theorists.  Indeed, Griffith et
al.\ (1998) show that a haze of photochemical aerosols might be
responsible for the relative smoothness of this part of the spectrum.
These aerosols may be activated by ultraviolet radiation from Gliese
229A (known to flare from time to time).  (A possible test of this
conjecture is that field brown dwarfs with effective temperatures near
900 K should not have these aerosols in their atmosphere since they
have no nearby source of incident ultraviolet radiation.  See Griffith
et al.\ 1998.)

This problem of whether silicate dusts, sulfides, hazes or even
polyacetylenes appear in the atmospheres of objects below the HBML
remains a subject of debate.  It is, however, extremely important
because dust can have a substantial effect upon the emergent spectrum.
The issue of whether a parent star can cause photochemical reactions
that greatly affect the spectra of its companions has implications for
planet and brown dwarf searches.  Indeed, the design of new searches
and instruments will need to take heed of this work.  In the case of
Gliese 229B, a spectrum in the 5 to 12 $\mu$m region might yield some
answers because incident ultraviolet radiation can also produce
certain organic molecules, such as C$_4$H$_2$, with spectral features
in the mid-infrared, as have been observed in spectra of Titan
(Griffith et al.\ 1998; Khlifi et al.\ 1997; Raulin and Bruston 1996).

\smallskip
\centerline{\psfig{file=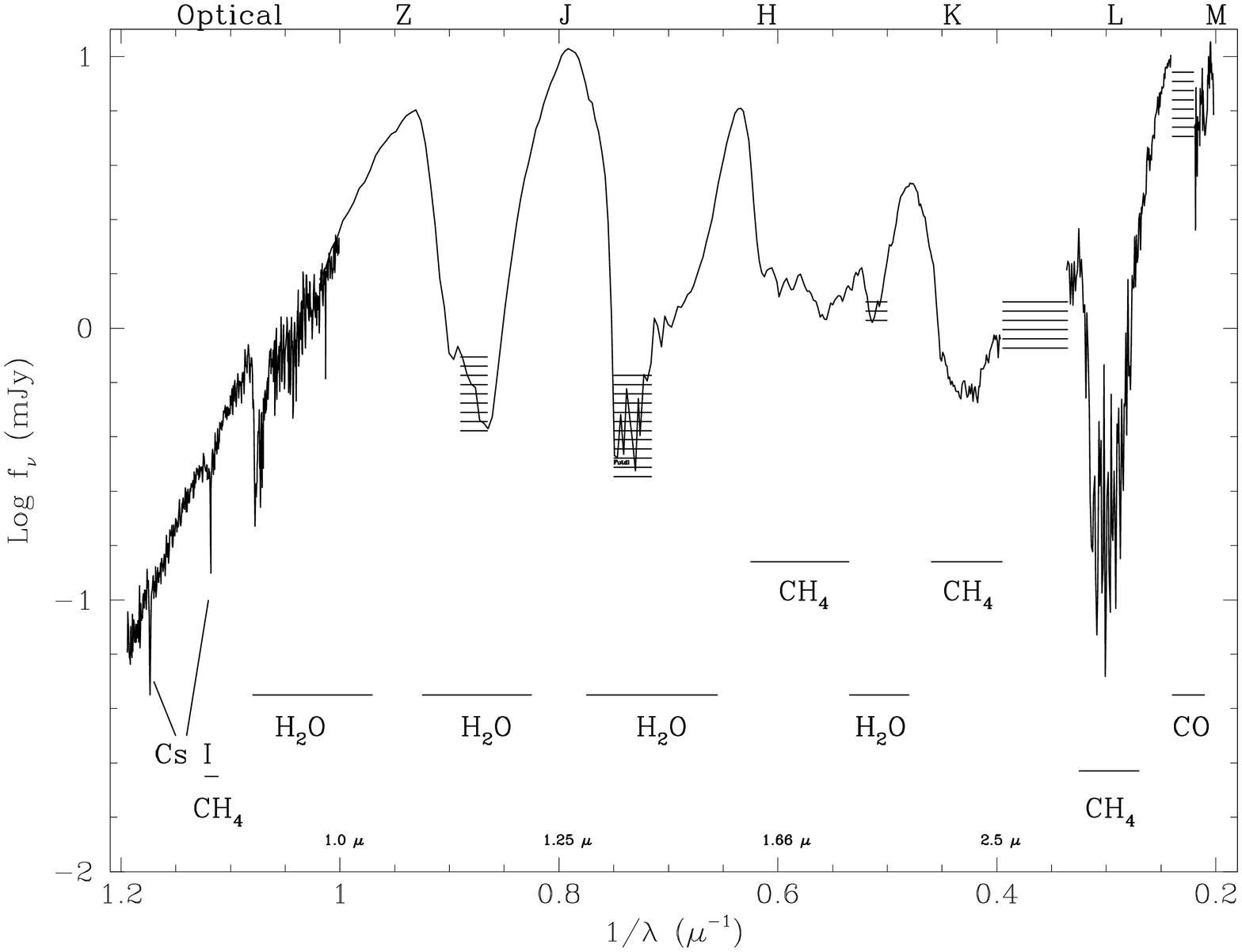,angle=270,width=4in}}

\noindent
{\caption{Fig.\ 7.\capskip The spectrum of Gliese 229B from 0.8 $\mu$m
to 5.0 $\mu$m.  The principal absorption features are, as indicated,
from water and methane, which is not present in stellar atmospheres.
Additional features due to cesium and carbon monoxide are also marked.
The relative smoothness of the 0.8 to 1.0 $\mu$m region indicates the
presence of an unpredicted continuum opacity source.  (From
Oppenheimer et al.\ 1998a.)}}

\smallskip

Noll et al.\ (1997) reported the detection of a feature due to carbon
monoxide in the 4 to 5 $\mu$m spectrum of Gliese 229B (Fig.\ 7).
Confirmed by Oppenheimer et al.\ (1998a), this feature shows that
carbon monoxide exists in non-equilibrium abundances.  At 900 K, the
balance between CO and CH$_4$ strongly favors CH$_4$.  Convection
must, therefore, dredge up appreciable quantities of CO from deeper,
hotter parts of the atmosphere.

However, Oppenheimer et al.\ (1998a) suggested that the convection
cannot be efficient even deeper where species such as VO and TiO ought
to be present because there are no spectral signatures of these
constituents.  VO and TiO are observed in all of the late M and L-type
dwarfs.  This may indicate that cool, old brown dwarfs are not fully
convective, but rather have an inner radiative zone below the outer
convective layer as the models of Burrows et al.\ (1998) show.

In Fig.\ 7 several absorption features due to neutral cesium are
indicated.  Oppenheimer et al.\ (1998a) argued that this should be the
last of the neutral atomic species present in atmospheres as one
proceeds toward lower and lower temperature.  Of all the known
elements, cesium has the lowest ionization potential, so at
temperatures well above Gliese 229B's it is ionized and its signature
is hidden in the extremely faint ultraviolet regions of the spectrum.
In addition, it, along with the other alkali metals, is less
refractory than the more familiar stellar atomic species (Al, Mg, Fe)
and so it survives in atomic form to lower temperatures.  Indeed, its
presence also in some of the L dwarf spectra (Kirkpatrick et al.\
1997) shows that it exists for effective temperatures from 1800 K to
below 900 K.  Oppenheimer et al.\ (1998a) suggest that neutral cesium,
along with the other alkali metals, be used as an indicator of
effective temperature for brown dwarfs.

Other than Gliese 229B, no other brown dwarfs have been confirmed as
companions of stars.  Oppenheimer et al.\ (1998b) have surveyed all of
the northern stars within 8 pc and found only one substellar
companion.  That survey was sensitive to objects up to 4 magnitudes
fainter than Gliese 229B with separations from the star between 3 and
30 arcsec.  The survey implies a star-brown dwarf binary frequency
of less than 1\%, although more than one specimen is needed to make
this statement significantly meaningful.  Several other searches are
underway, however, including that of Krist et al.\ (1998) with the
Space Telescope.  It seems clear from the lack of numerous brown dwarf
companions that hundreds or thousands of stars must be surveyed before
a substantial population of these elusive objects can be studied in
detail.

\subsection{\bf Acknowledgments}
SRK would like to particularly thank T. Nakajima for getting him
interested in brown dwarfs during his postdoctoral stay at Caltech.
We would also like to thank M. Zapatero-Osorio, E. Mart\'{\i}n and
R. Rebolo for being so helpful with figures, A. Burrows for ever
useful and engaging discussions and the use of several of his figures,
G. Basri, E. Mart\'{\i}n, B. Brandl, G. Vasisht and K. Adelberger for
thorough comments on the draft.  We also thank the NSF and NASA for
support of our brown dwarf research.

\vfill\eject
\null

\vskip .5in
\centerline{\bf REFERENCES}
\vskip .25in

\ref{Alcock, C. et al.\ 1998. EROS and MACHO Combined Limits on 
Planetary-mass Dark Matter in the Galactic Halo.  {\refit
Astrophys.\ J.\ Lett.\/} 499:L9--L12.}

\ref{Allard, F., Hauschildt, P. H., Baraffe, I. and Chabrier,
G. 1996.  Synthetic Spectra and Mass Determination of the Brown Dwarf
Gliese 229B.  {\refit Astrophys.\ J.\ Lett.\/} 465:L123--127.}

\ref{Allard, F., Hauschildt, P. H., Alexander, D. R. and Starrfield,
S. 1997.  Model Atmospheres of Very Low-Mass Stars and Brown Dwarfs.
{\refit Ann.\ Rev.\ Astron.\ Astrophys.\/} 35:137--177.}

\ref{Baraffe, I., Chabrier, G., Allard, F., and Hauschildt,
P. H. 1995.  New Evolutionary Tracks for Very Low-Mass Stars. {\refit
Astrophys. J. Lett.\/} 446:L35--L38.}

\ref{Basri, G. 1998.  The Lithium Test for Young Brown Dwarfs. In {\refit
Brown Dwarfs and Extrasolar Planets}, R. Rebolo, E. L. Mart\'{\i}n and
M. Zapatero-Osorio, eds. (San Francisco: Astronomical Society of the
Pacific, Conference Series) 134:394--404.}

\ref{Basri, G., Marcy, G. and Graham, J.  1996.  Lithium in Brown Dwarf
Candidates: The Mass and Age of the Faintest Pleiades Stars. {\refit
Astrophys.\ J.\/} 458:600--609.}

\ref{Basri, G. and Marcy, G. 1997. Early Hints on the Substellar Mass
Function.  In {\refit Star Formation Near and Far}, S. S. Holt and
L. G. Mundy, eds. (New York: AIP Press) 228--240.}

\ref{Basri, G. and Mart\'{\i}n, E. L.  1998.  PPL 15: the First Binary
Brown Dwarf System?  In {\refit Brown Dwarfs and Extrasolar Planets},
R. Rebolo, E. L. Mart\'{\i}n and M. Zapatero-Osorio, eds.\ (San
Francisco: Astronomical Society of the Pacific, Conference Series)
134:284--287.}

\ref{Basri, G. and Mart\'{\i}n, E.L.  1999. The Mass and Age of 
Very Low-Mass Members of the Open Cluster $\alpha$ Persei. {\refit
Astrophys.\ J.\/} in press.}

\ref{Becklin, E. E. and Zuckerman, B.  1988. A Low-temperature
Companion to a White Dwarf Star.  {\refit Nature} 336:656--658.}

\ref{Beichman, C. A. 1987.  The IRAS View of the Galaxy and the
Solar-System. {\refit Ann.\ Rev.\ Astron.\ Astrophys.\/} 25:521--563.}


\ref{Bildsten, L., Brown, E. F., Matzner C. D., Ushomirsky G.  1997.
Lith-ium Depletion in Fully Convective Pre-Main-Sequence Stars. {\refit
Astrophys.\ J.\/} 482:442--447.}

\ref{Black, D. C.  1997. Possible Observational Criteria for
Distinguishing Brown Dwarfs from Planets. {\refit
Astrophys.\ J.\ Lett.\/} 490:L171--L174.}

\ref{Bouvier, J., Stauffer, J., Mart\'{\i}n, E., Barrado y
Navascu\'es, D., Wallace, B., and Bejar, V.  1998.  Brown Dwarfs and Very
Low-Mass Stars in the Pleiades Cluster:  A Deep Wide-Field Imaging Survey.
{\refit Astron.\ Astrophys.\/} 335:183--198.}

\ref{Burkert, A. and Bodenheimer, P. 1993.  Multiple Fragmentation in 
Collapsing Protostars.  {\refit Mon.\ Not.\ Roy.\ Astron.\ Soc.\/}
264:798--806.}

\ref{Burrows, A. and Liebert, J. 1993.  The Science of Brown
Dwarfs. {\refit Rev.\ Mod.\ Phys.\/} 65:301--336.}

\ref{Burrows, A., Marley M., Hubbard W. B., Lunine J. I., Guillot T.,
Saumon D., Freedman R., Sudarsky D. and Sharp C.  1997.  A Nongray
Theory of Extrasolar Giant Planets and Brown Dwarfs. {\refit
Astrophys.\ J.\/} 491:856--875.}

\ref{Burrows, A. and Sharp, C. M. 1998. Chemical Equilibrium
Abundances in Brown Dwarf and Extrasolar Giant Planet
Atmospheres. {\refit Astrophys.\ J.\/}, in press.}

\ref{Chabrier, G. and Baraffe, I. 1997. Structure and evolution of 
low-mass stars. {\refit Astron.\ Astrophys.\/} 327:1039--1053.}

\ref{Comeron, F., Rieke, G., Claes, P., and Torra, J.  1998.  ISO Observations
of Candidate Young Brown Dwarfs.  {\refit Astron.\ Astrophys.\/}
335:522--532.}

\ref{D'Antona, F. and Mazzitelli, I. 1994. New Pre-Main Sequence Tracks 
for $M < 2.5 M_\odot$ as Tests of Opacities and Convection
Model. {\refit Astrophys.\ J.\ Suppl.\ Ser.\/} 90:467--500.}

\ref{D'Antona, F. 1995. Luminosity Functions and the Mass
Function. In {\refit The Bottom of the Main Sequence --- and Beyond},
C. G. Tinney, ed. (Berlin: Springer) pp. 13--23.}

\ref{Delfosse, X., Tinney, C., Forveille, T., Epchtein, N.,
Bertin, E. Borsenberger, J., Copet, E., de Batz, B., Fouque, P.,
Kimeswenger, S., LeBertre, T., Lacombe, F., Rouan, D., and Tiphene,
D.  1997.  Field Brown Dwarfs Found by DENIS.  {\refit Astron.\
Astrophys.\/} 327:L25--L28.}

\ref{Fegley, B. Jr. and Lodders, K. 1994.  Chemical Models of the Deep
Atmospheres of Jupiter and Saturn. {\refit Icarus} 110:117--154.}

\ref{Forrest, W. J., Skrutskie, M. F. and Shure, M. 1988.  A Possible
Brown Dwarf Companion to Gliese 569. {\refit Astrophys.\ J.\ Lett.\/}
330:L119--L123.} 

\ref{Geballe, T. R., Kulkarni, S. R., Woodward, C. E. and Sloan,
G. C.  1996.  The Near-Infrared Spectrum of the Brown Dwarf Gliese
229B. {\refit Astrophys.\ J.\ Lett.\/} 467:L101--L104.}

\ref{Griffith, , C. A., Yelle, R. V. and Marley, M. 1998.  The Dusty
Atmosphere of Gliese 229B. {\refit Science}, in press.}

\ref{Golimowski, D. A., Burrows, C. J., Kulkarni, S. R., Oppenheimer,
B. R. and Brukardt, R. A. 1998.  Wide Field Planetary Camera 2
Observations of the Brown Dwarf Gliese 229B: Optical Colors and
Orbital Motion. {\refit Astron.\ J.\/} 115:2579--2586.}

\ref{Hambly, N., Hawkins, M. R. S. and Jameson, R. F.  1993.
Very Low Mass Members of the Pleiades. {\refit Astron.\ Astrophys.\/}
100:607--640.}

\ref{Henry, T. J. and McCarthy, D. W. Jr.  1990.  A Systematic Search
for Brown Dwarfs Orbiting Nearby Stars. {\refit Astrophys.\ J.\/}
350:334--347.}

\ref{Henry, T. J. and McCarthy, D. W. Jr.  1993.  The Mass-Luminosity
Relation for Stars of Mass 1.0 to 0.08 $M_\odot$. {\refit Astron.\
J.\/} 106:773--789.}

\ref{Henry. T. J., McCarthy, D. W. Jr., Freeman, J. and Christou,
J. C. 1992.  Nearby Solar-Type Star with a Low-Mass Companion---New
Sensitivity Limits Reached Using Speckle Imaging. {\refit Astron.\
J.\/} 103:1369--1373.}



\ref{Hodgkin, S. and Jameson, R. F. 1997.  Brown Dwarfs.  {\refit
Cont.\ Phys.\/} 38:395--407.}

\ref{Irwin, M., McMahon, R., and Hazard, C.  1991.  APM Optical Surveys
for High Redshift Quasars.  In {\refit The Space Distribution of
Quasars}, D. Crampton, ed. (San Francisco: Astronomical Society of the
Pacific, Conference Series) 21:117--126.}

\ref{Jameson, R. F. and Skillen, I. 1989.  A Search for Low-Mass Stars
and Brown Dwarfs in the Pleiades. {\refit Mon.\ Not.\ Roy.\ Astron.\
Soc.\/} 239:247--253.}

\ref{Jones, H. R. A., Longmore, A. J., Jameson, R. F., Mountain, M.
1994.  An Infrared Spectral Sequence for M Dwarfs. {\refit Mon.\ Not.\
Roy.\ Astron.\ Soc.\/} 267:413--423.}

\ref{Jones, H. R. A. and Tsuji, T. 1997. Spectral Evidence for Dust in
Late-Type M Dwarfs.  {\refit Astrophys.\ J.\ Lett.\/} 48:L39--L42.}

\ref{Kafatos, M. C., Harrington, R. S. and Maran,
S. P. eds. 1986. {\refit Astrophysics of Brown Dwarfs} (New York:
Cambridge University Press).}

\ref{Kerins, E. and Wyn Evans, N. 1998  Microlensing Halo Models with 
Abundant Brown Dwarfs. {\refit Astrophys.\ J.\ Lett.\/} 503:L75--L78.}

\ref{Khlifi, M., Paillous P., Bruston P., Guillemin J. C., Benilan
Y., Daoudi A. and Raulin F.  1997.  Gas Infrared Spectra, Assignments,
and Absolute IR Band Intensities of C$_4$N$_2$ in the 250-3500 cm(-1)
Region: Implications for Titan's Stratosphere. {\refit Spectrochimica
Acta A} 53:707--712.}

\ref{Kirkpatrick, J.D.  1997. Spectroscopic Properties of Ultra-Cool
Dwarfs and Brown Dwarfs.  In {\refit Brown Dwarfs and Extrasolar
Planets}, R. Rebolo, E. Mart\'{\i}n and M. Zapatero-Osorio, eds.\ (San
Francisco: Astronomical Society of the Pacific, Conference Series)
134:405--415. }

\ref{Kirkpatrick, J.D., Henry, T. and Irwin, M.  1997.  Ultra-Cool M Dwarfs
Discovered by QSO Surveys.  I: The APM Objects.  {\refit Astron.\
J.\/} 113:1421--1428.}

\ref{Kirkpatrick, J. D., Allard, F., Bida, T., Zuckerman, B., Becklin, E. E., 
Chabrier, G. and Baraffe, I. 1998.  An Improved Optical Spectrum and
New Model Fits of the Likely Brown Dwarf GD 165B. {\refit Astrophys.\
J.\/} submitted.}

\ref{Krasnopolsky, V. A., Bjoraker, G. L., Mumma, M. J. and 
Jennings, D. E.  1997.  High-Resolution Spectroscopy of Mars at 3.7
and 8 $\mu$m: A Sensitive Search for H$_2$O$_2$, H$_2$CO, HCl, and
CH$_4$, and Detection of HDO. {\refit J.\ Geophys.\ Res.\/}
102:6525--6534.}

\ref{Krist, J. E., Golimowski, D. A., Schroeder, D. J. and Henry,
T. J. 1998. Characterization and Subtraction of Well-Exposed
HST/NICMOS Camera 2 Point Spread Functions for a Survey of Very
Low-Mass Companions to Nearby Stars. {\refit Publ.\ Astron.\ Soc.\
Pac.\/} in press.}

\ref{Kulkarni, S. R. 1997.  Brown Dwarfs: A Possible Missing Link
Between Stars and Planets. {\refit Science} 276:1350--1354.}

\ref{Kumar, S. S. 1963. The Structure of Stars of Very Low Mass.  
{\refit Astrophys.\ J.\/} 137:1121-1126.}


\ref{Leggett, S., and Hawkins, M.R.  1988.  The Infrared Luminosity Function
for Low-Mass Stars.  {\refit Mon.\ Not.\ Roy.\ Astron.\ Soc.\/}
234:1065--1090.}

\ref{Leggett, S., Harris, H., and Dahn, C.  1994.  Low-Mass Stars in the
Central Region of the Hyades Cluster.  {\refit Astron.\ J.\/}
108:944--963.}

\ref{Luhman, K. L., Liebert, J., and Rieke, G. H.  1997.  Spectroscopy 
of a Young Brown Dwarf in the Rho Ophiuchi Cluster.  {\refit
Astrophys.\ J.\ Lett.\/} 489:L165--L168.}

\ref{Luhman, K. L., Rieke, G. H., Lada, C., and Lada, E.  1998. 
Spectroscopy of Low-Mass Members of IC348. {\refit Astrophys.\ J.\/}
508: submitted.}

\ref{Luyten, W. and Kowal, C. 1975.  {\refit Proper Motion Survey with the
48 Inch Schmidt Telescope} (Minneapolis: Observatory of the University
of Minnesota).}

\ref{Magazz\`u, A., Rebolo, R., Zapatero Osorio, M., and Mart\'{\i}n, E.,
and Hodgkin, S.  1998.  A Brown Dwarf Candidate in the Praesepe Open
Cluster.  {\refit Astrophys.\ J.\ Lett.\/} 497:L47--L51.}

\ref{Marcy, G., Basri, G. and Graham, J.  1994.  A Search for 
Lithium in Pleiades Brown Dwarf Candidates Using the Keck HIRES
Echelle.  {\refit Astrophys.\ J.\ Lett.\/} 428:L57--L60.}

\ref{Marley, M. S., Saumon, D., Guillot, T., Freedman, R. S., Hubbard,
W. B., Burrows, A. and Lunine, J. I.  1996.  Atmospheric,
Evolutionary, and Spectral Models of the Brown Dwarf Gliese 229 B.
{\refit Science.\/} 272:1919--1921.}

\ref{Mart\'{\i}n, E. L., Rebolo, R. and Magazz\`u, A.  1994.
Constraints to the Masses of Brown Dwarf Candidates from the Lithium
Test. {\refit Astrophys.\ J.\/} 436:262--269.}

\ref{Mart\'{\i}n, E., Basri, G., Delfosse, X., and Forveille, T.  1997.
Keck HIRES Spectra of the Brown Dwarf DENIS-PJ1228.2-1547.
{\refit Astron.\ Astrophys.\/} 327:L29--L32.}

\ref{Mart\`{\i}n, E. L., Basri, G., Gallegos, J. E., Rebolo, R.,
Zapatero-Osorio, M. R. and Bejar, V. J. S. 1998.  A New Pleiades
Member at the Lithium Substellar Boundary. {\refit Astrophys.\ J.\
Lett.\/} 499:L61--L64.}

\ref{Matthews, K., Nakajima, T., Kulkarni, S. R. 1996.  Spectral
Energy Distribution  and Bolometric Luminosity of the Cool Brown Dwarf
Gliese 229B.  {\refit Astron.\ J.\/} 112:1678--1682.}

\ref{McCarthy, D. W., Jr., Probst, R. G. and Low, F. J.  1985.
Infrared Detection of a Close Cool Companion to van Biesbroeck
8. {\refit Astrophys.\ J.\ Lett.\/} 290:L9--L13.}

\ref{Meynet, G., Mermilliod, J.-C. and Maeder, A. 1993. New Dating of Galactic
Open Clusters. {\refit Astron.\ Astrophys.\ Suppl.\ Ser.\/} 98:477--504.}

\ref{Nakajima, T., Durrance, S. T., Golimowski, D. A. and Kulkarni,
S. R.  1994.  A Coronagraphic Search for Brown Dwarfs around Nearby
Stars.  {\refit Astrophys.\ J.\/} 428:797--804.}

\ref{Nakajima, T., Oppenheimer, B. R., Kulkarni, S. R., Golimowski,
D. A., Matthews, K. and Durrance, S. T.  1995. Discovery of a Cool Brown
Dwarf. {\refit Nature.\/} 378:463--465.}

\ref{Noll, K. S. 1993. Spectroscopy of Outer Solar System Atmospheres
from 2.5 to 7.0 Microns.  In {\refit Astronomical Infrared
Spectroscopy: Future Observational Directions}, S. Kwok, ed. (San
Francisco: Astronomical Society of the Pacific, Conference Series)
41:29--40.} 

\ref{Noll, K. S., Geballe, T. R, and Marley, M. S. 1997.  Detection of 
Abundant Carbon Monoxide in the Brown Dwarf Gliese 229B. {\refit
Astrophys.\ J.\ Lett.\/} 489:L87--L90.}

\ref{Oppenheimer, B. R., Kulkarni, S. R., Matthews, K. and Nakajima,
T.  1995.  Infrared Spectrum of the Cool Brown Dwarf GL 229B. {\refit
Science.\/} 270:1478--1479.}

\ref{Oppenheimer, B. R., Kulkarni, S. R., Matthews, K. and van
Kerkwijc, M. H.  1998a.  The Spectrum of the Brown Dwarf Gliese 229B.
{\refit Astrophys.\ J.\/}  502:932--943.}

\ref{Oppenheimer, B. R., Golimowski, D. A., Kulkarni, S. R. and 
Matthews, K. 1998b.  An Optical Coronagraphic and Near IR Imaging
Survey of Stars for Faint Companions. In preparation.}

\ref{Pavlenko, Y. V., 1998.  Lithium Lines in the Spectra of Late M
Dwarfs.  {\refit Astron.\ Rep.\/} 42:501--507.}

\ref{Pinfield, D., Hodgkin, S., Jameson, R., Cossburn, M. and
von Hippel, T.  1997.  Brown Dwarf Candidates in Praesepe.  {\refit
Mon.\ Not.\ Roy.\ Astron.\ Soc.\/} 287:180--188.}

\ref{Pinsonneault, M., Stauffer, J., Soderblom, D., King, J.,
and Hanson, R.  1998.  The Problem of Hipparcos Distances to Open
Clusters: I. Constraints from Multicolor Main Sequence Fitting.
{\refit Astrophys.\ J.\/} in press.}

\ref{Pound, M. W. and Blitz, L. 1995.  Proto-Brown Dwarfs 2: Results
in the Ophiuchus and Taurus Molecular Clouds. {\refit Astrophys.\
J.\/} 444:270--287.}

\ref{Probst, R. G. 1983. An Infrared Search for Very Low Mass Stars:
The Luminosity Function.  {\refit Astrophys.\ J.\/} 274:237--244.}

\ref{Raboud, D., and Mermilliod, J.-C.  1998.  Investigation of the Pleiadees
Cluster. IV. The Radial Structure.  {\refit Astron.\ Astrophys.\/}
329:101--114.}

\ref{Raulin, F. and Bruston, P. 1996. Photochemical Growing of 
Complex Organics in Planetary Atmospheres. {\refit Life Sci.\/} 18:41--49.}

\ref{Rebolo, R.,  Mart\'{\i}n, E., and Magazz\`u, A.  1992.  Spectroscopy of a
Brown Dwarf Candidate in the $\alpha$ Persei Open Cluster.  {\refit
Astrophys.\ J.\/} 389:L83--L87.}

\ref{Rebolo, R., Zapatero Osorio, M. and Mart\'{\i}n, E.  1995.
Discovery of a Brown Dwarf in the Pleiades Star Cluster.  {\refit
Nature} 377:129--131.}

\ref{Rebolo, R., Mart\'{\i}n, E. L., Basri, G., Marcy, G. and
  Zapatero Osorio, M.R.  1996.  Brown Dwarfs in the Pleiades Cluster
Confirmed by the Lithium Test.  {\refit Astrophys.\ J.\ Lett.\/}
469:L53--L57.}

\ref{Rebolo, R., Mart\'{\i}n, E. L. and Zapatero-Osorio,
M. R. eds. 1998. {\refit Brown Dwarfs and Extrasolar Planets} (San
Francisco: Astronomical Society of the Pacific) Conference Series
Vol. 134.}

\ref{Rieke, G. H. and Rieke, M. J. 1990.  Possible Substellar Objects 
in the Rho Ophiuchi Cloud.  {\refit Astrophys.\ J.\ Lett.\/}
362:L21--L24.}

\ref{Ruiz, M. T., Leggett, S. K., and Allard, F.  1997.  Kelu-1: A 
Free-Floating Brown Dwarf in the Solar Neighborhood. {\refit
Astrophys.\ J.\ Lett.\/} 491:L107--L111.}

\ref{Schultz, A. B. et al. 1998. First Results from the Space
Telescope Imaging Spectrograph: Optical Spectra of Gliese 229B.
{\refit Astrophys.\ J.\ Lett.\/} 492:L181--L182.}

\ref{Shu, F. H., Adams, F. C. and Lizano, S. 1987. Star Formation in 
Molecular Clouds: Observation and Theory. {\refit Ann.\ Rev.\ Astron.\
Astrophys.\/} 25:23--81.}

\ref{Simons, D. A., Henry, T. J., and Kirkpatrick, J. D.  1996.  The
Solar Neighborhood. III.  A Near-Infrared Search for Widely Separated
Low-Mass Binaries.  {\refit Astron.\ J.\/} 112:2238--2247.}

\ref{Stauffer, J., Hamilton, D., Probst, R., Rieke, G. and
Mateo, M.  1989.  Possible Pleiades Members with M $\sim$ 0.07 Solar
Mass: Identification of Brown Dwarfs of Known Age, Distance and
Metallicity.  {\refit Astrophys.\ J.\ Lett.\/} 344:L21--L25.}

\ref{Stauffer, J., Liebert, J., Giampapa, M., MacIntosh, B., Reid, I.N.,
 and Hamilton, D.  1994.  Radial Velocities of Very Low Mass Stars and
Candidate Brown Dwarf Members of the Hyades and Pleiades.  {\refit
Astron.\ J.\/} 108:160--174.}

\ref{Stauffer, J., Schultz, G., and Kirkpatrick, J.D.  1998.  Keck Spectra
of Pleiades Brown Dwarf Candidates and a Precise Determination of the
Lithium Depletion Edge in the Pleiades.  {\refit Astrophys.\ J.\
Lett.\/} 499:L199--L203.}

\ref{Stringfellow, G.  1991.  Brown Dwarfs in Young Stellar Clusters.
{\refit Astrophys.\ J.\ Lett.\/} 375:L21--L25.}

\ref{Tartar, J. C. 1986. {\refit Astrophysics of Brown Dwarfs},
M. C. Kafatos, R. S. Harrington and S. P. Maran, eds.\ (Cambridge:
Cambridge University Press).}

\ref{Thackrah, A., Jones, H., and Hawkins, M.  1997.  Lithium 
Detection in a Field Brown Dwarf Candidate.  {\refit Mon.\ Not.\ Roy.\
Astron.\ Soc.\/} 284:507--512.}

\ref{Tinney, C. G.  1995.  ed. {\it The Bottom of the Main Sequence 
--- and Beyond.} (Berlin: Springer).}

\ref{Tinney, C. G.  1998.  The Intermediate Age Brown Dwarf LP 944-20.
{\refit Mon.\ Not.\ Roy.\ Astron.\ Soc.\/} 296:L42--L44.}

\ref{Tinney, C. G., Delfosse, X. and Forveille, T. 1997.  DENIS-P 
J1228.2-1547 --- A New Bench Mark Brown Dwarf.  {\refit Astrophys.\ J.\
Lett.\/} 490:L95--L98.}

\ref{Toth, R. A. 1997.  Measurements of HDO between 4719 and 5843
cm(-1).  {\refit J.\ Mol.\ Spect.\/} 186:276--292.}

\ref{Tsuji, T. 1964. {\refit Ann.\ Tokyo Astron.\ Obs.\ Ser.\ II} 9:1--26.}

\ref{Tsuji, T., Ohnaka, K., Aoki, W. and Nakajima, T.  1996.
Evolution of Dusty Photospheres through Red to Brown Dwarfs: How Dust
Forms in Very Low-Mass Objects. {\refit Astron.\ Astrophys.\ Lett.\/}
308:L29--L32.}

\ref{Ventura, P., Zeppieri, A., Mazzitelli, I., D'Antona,
F. 1998. Pre-Main Sequence Lithium Burning: The Quest for a New
Structural Parameter. {\refit Astron.\ Astrophys.\/} 331:1011--1021.}

\ref{ Wilking, B. A., Green, T. P. and Meyer, M. R.,
1998. Spectroscopy of Brown Dwarf Candidates in the $\rho$ Oph
Molecular Core.  {\refit Astron.\ J.\/} submitted.}

\ref{Zapatero-Osorio, M. R., 1997. New Brown Dwarfs in the Pleiades
Cluster. {\refit Astrophys.\ J.\ Lett.\/} 491:L81--L84.}

\ref{Zuckerman, B. and Becklin, E. E. 1992.  Companions to White
Dwarfs --- Very Low-Mass Stars and the Brown Dwarf Candidate GD 165B.
{\refit Astrophys.\ J.\/} 386:260--264.}

\bye